\documentclass[useAMS,usenatbib,referee]{mn2e}
\def\lsim{\lower.5ex\hbox{$\; \buildrel < \over \sim \;$}}
\def\gsim{\lower.5ex\hbox{$\; \buildrel > \over \sim \;$}}
\def\be{\begin{equation}}
\def\ee{\end{equation}}

\def\md{\dot{\cal M}}

\def\vt{v_{\infty}}
\def\gmt{\gamma_{\infty}}
\def\msol{{\rm M}_\odot}
\def\mbh{M_{\rm B}}
\def\bc{\begin{center}}
\def\ec{\end{center}}
\def\eg{{\it e.g.,}}
\def\etal{{\em et al.}}
\def\ie{{\em i.e.,}}
\def\ep{{{\rm e}^--{\rm p}^+}}
\def\eel{{{\rm e}^--{\rm e}^+}}

\def\grb{{\rm GRB}}
\def\rg{r_{\rm g}}
\def\rs{r_{\rm s}}
\def\tg{t_{\rm g}}
\def\xsh{x_{\rm sh}}
\def\bsh{B_{\rm sh}}
\def\Lj{L_{\rm j}}
\def\rsh{r_{\rm sh}}
\include{netbib.sty}

\usepackage{graphicx}
\usepackage{color}
\title[Thermally driven jets and internal shocks]
{General relativistic study of astrophysical jets with internal shocks}
\author[Vyas \& Chattopadhyay]
{Mukesh K. Vyas$^{1}$, Indranil Chattopadhyay$^{1}$ \thanks{Email:mukesh.vyas@aries.res.in (MKV);
		indra@aries.res.in (IC)}\\
$^{1}$Aryabhatta Research Institute of Observational Sciences 
(ARIES), Manora Peak, Nainital-263002, India\\
}
\begin{document}
\date{}
\maketitle
\label{firstpage}

\begin{abstract}
We explore the possibility of formation of steady internal shocks in jets around black holes.
We consider a fluid described by a relativistic equation of state, flowing about the
axis of symmetry ($\theta=0$) in a Schwarzschild metric.
We use two models for the jet geometry, (i) a conical geometry and (ii) a geometry
with non-conical cross-section.
Jet with conical geometry is smooth flow. While the jet with non-conical cross
section undergoes multiple sonic point and even standing shock.
The jet shock becomes stronger, as the shock location is situated further from the
central black hole. Jets with very high energy and very low energy do not harbour
shocks, but jets with intermediate energies do harbour shocks. 
One advantage of these shocks, as opposed to shocks mediated by external
medium is that, these shocks have no effect on the jet terminal speed, 
but may act as possible sites for particle acceleration.
Typically, a jet with energy $1.8~c^2$, will achieve a terminal speed
of $v_\infty=0.813c$ for jet with any geometry. But for a jet of non-conical cross-section
for which the length scale of the inner torus of the accretion disc is $40\rg$, then
in addition, a steady shock will form at $\rsh \sim 7.5\rg$ and compression ratio of
$R\sim 2.7$. Moreover, electron-proton jet seems to harbour the strongest shock.
We discuss possible consequences of such a scenario.
\end{abstract}


\section {Introduction}
The denomination relativistic jets was first used in the extra-galactic context by \cite{bm54},
while observing knots in optical waveband 
in galaxy M87. However, it is after the 
advent of radio astronomy, that the 
relativistic jets have been recognized 
as a very common phenomenon in astrophysics and is 
associated with various kinds of 
astrophysical objects such as active galactic nuclei 
(AGN e.g., M87), gamma ray bursts (\grb), young stellar objects 
(YSO e.g., HH 30, HH 34), X-ray binaries (\eg SS433, Cyg X-3, 
GRS 1915+105, GRO 1655-40) etc. 
In this paper, we discuss about jets which are associated around
X-ray binaries like GRS1915+105 \citep{mr94} and AGNs like 3C273, 3C345 \citep{zcu95}, M87 \citep{b93}.
Since jets are unlikely to be ejected from the surface of compact objects, therefore, it has to
originate from accreting matter.

High luminosity and broad band spectra from AGNs, favour the model of
accretion of matter and energy on to a super massive ($10^{6-9}\msol$) black hole as the prime mover.
The same line of reasoning also
led to the conclusion that X-ray binaries harbour neutron stars or stellar mass
($\lsim 10\msol$) black holes
at the centre, while the secondary star feeds the compact object.
Many of the black hole (BH) X-ray binaries are observed to go through a stage, in which
relativistic twin jets are ejected
and resemble scaled down version of AGN or quasar jets; as a result,
these sources have been coined micro-quasars \citep{mrcpl92}. 
Accretion of matter on to a compact object explains the luminosities and spectra
of AGNs and micro-quasars, but can accretion be linked with the formation of jets?
Interestingly, simultaneous 
radio and X-ray observations of micro-quasars show a very 
strong
correlation between the spectral states of the accretion 
disc and the associated jet states \citep{gfp03,fgr10,rsfp10}, 
which reaffirms the fact that jets do originate from the accretion 
disc. Although this type of spectral state changes and the related change in jet states
have not been observed for AGNs, however, the very fact that the timescales of AGNs and
micro-quasars can be scaled by the central mass \citep{mkkf06}, 
we expect similar correlation of spectral and jet states
in AGNs too. Working with the central idea that jets are launched from accretion discs, a natural question arises, i.e., which part of the disc is responsible for jet generation.
Recent observations have shown that jets 
originate from a region which is less than 100 Schwarzschild radius
($\rs$) around the unresolved central object \citep{jbl99,detal12}, 
which implies 
that the entire disc may not participate in formation of jets, but only 
the central region of the disc is responsible. 
Again by invoking the similarity of AGNs and micro-quasars \citep{mkkf06}, one may conclude that the jet
originates from a region close to the central compact object for micro-quasars too.
This also finds indirect support from observations of various micro quasars.
Observations showed that the jet activity starts when the object is in
low-hard state (LHS i. e., when the disc emission maximizes in hard, or,
high energy X-rays but the over all
luminosity is low). The jet strength increases as accretion disc becomes luminous in the intermediate hard states (IHS) and eventually with a surge in disc luminosity and
after relativistic ejections, the accretion disc goes into
high soft state (HSS i. e., disc emission maximizes in soft X-rays but luminous).
No jet activity has been observed in HSS. This cycle is repeated as some kind of hysteresis and is known
as hardness-intensity-diagram or HID \citep{fbg04}. This indirectly suggests that the component
of the disc which emits hard X-rays, or a compact corona, may also be responsible for jet activity. Since hard X-rays
are emitted by hot electron clouds closer to the compact object, it indirectly points out that the jet
originates from a region close to the BH even for micro-quasars. 

Earliest model of accretion disc was the Keplerian disc or KD \citep[disc with Keplerian angular momentum distribution and optically thick along the radial direction, see][]{ss73,nt73}. KD explained
the thermal part of the BH accretion disc spectra, but could not explain the non-thermal part of it.
This deficiency in the KD, prompted the emergence of many accretion disc models like thick disc \citep{pw80}, advection dominated accretion flow
\citep{nkh97}, advective discs \citep{lt80,f87a,c89,cc11}. Whatever, may be the accretion disc models,
observation of
strong correlation of the X-ray data (arising from convergent flow and therefore accretion)
and radio data (originating from the outflowing jets) in microquasars, and the associated timing properties,
also imposed some constraints on the disc-jet system.
In micro quasars it was observed that the hard photons oscillate in a quasi periodic manner
and which evolves from low values in LHS to high values in the intermediate states, and disappears in the HSS. This suggests that a compact corona is favourable,
than an extended one, since it would be easier to oscillate a compact corona than an extended one.
It was also suggested that the extended corona cannot explain the spectra
of X-ray binaries in the LHS \citep{dwmb97,getal97}. So from observations, one can summarize about three aspects
of jet and accretion discs: they are (i) direct observation of inner region of M87 jet shows, the jet base is close to the central object; (ii) corona emits non-thermal emission, and the HSS state has very weak or no signature
of corona, as well as, the existence of jets and (iii) the corona is also compact in size.
So it is quite possible that, corona is probably the base of the jet, or, a significant part of the jet base.   

Unfortunately, the accretion disc around a BH has not been resolved and
only the jet has actually been observed, therefore, studying
the visible part of the jet is also an integral part of reconstructing the entire picture, and hence
jets, especially the AGN jets are intensely investigated. 
Generally, it is assumed that jets from AGNs are ultra-relativistic with 	
terminal Lorentz factors $\gmt \gsim 10$. However, with the discovery of more AGNs
and increasingly precise observations, these facts are slowly being challenged
now.
On one hand, for BL Lac PKS 2155-304 the estimated terminal Lorentz factor is truly
relativistic i. e., easily $\gmt \gsim 10$ \citep{ah07},
on the other hand,
spectral fitting of NGC 4051 requires a more moderate range of Lorentz factors \citep{mmmk11}.
Infact, for FRI type jet 3C31, which has well resolved jet and counter jet observations,
the jet terminal speed is $\vt \sim 0.8c \rightarrow 0.85c$ ($c$ the speed of light)
and then slowing
down to $\sim 0.2c$ at few kpc due to entrainment with ambient medium \citep{lb02}.
This implies that jets of AGNs comes at a variety of strength, length and terminal speeds ($\vt$),
somewhat similar to those around microquasars. For example in microquasars, SS433 jet shows a quasi-steady jet with $\vt\sim 0.26c$ \citep{m84},
while GRS1915+105 jet is truly relativistic \citep{mr94}. Not only that, terminal speeds estimated
for a single micro-quasar may vary in different outbursts \citep{metal12}.
In other words, astrophysical jets around compact objects are relativistic, but the terminal
speed may vary from being mildly relativistic to ultra-relativistic. Therefore, acceleration mechanism
must be multi-staged and may be result of many accelerating processes like magnetic fields and
radiation driving \citep{sw81,fr85,f96,f00,fth01,vt99,cc02a,cc02b,cdc04,c05,kcm14,vkmc15}.

Estimation of bulk jet speeds from AGN jets are mostly inferred from complicated observational data.
Often, the presence of bright jet in comparison to dim counter jet, are believed to constrain the
bulk speed of the jet \citep{wa97}. It has also been noted above that
bulk speed required to fit the observed data, also gives us an estimate of the bulk speed of the jet
\citep{lb02}.
 Many of these jets have knots and hot spots, which are regions
of enhanced brightness. Some
of these knots exhibit superluminal speeds \citep{bsm99}, which give us an estimate of the
bulk speed of the knots and the underlying jet speed.
These knots and hot spots are thought to arise due to the presence of shocks
in the jet beam, due to its interaction with the ambient medium. These shocks then create
high energy electrons by shock acceleration and produce non-thermal,
high energy photons. By analyzing multi-wavelength
data of the M87 jet, \citet{pw05} concluded that external shocks fit this general idea pretty well.
From the accumulated knowledge of hydrodynamic simulations, we know that these
shocks form as a result of interaction with the ambient medium \citep{mmfim97} and therefore
they form at large distances from the central object. However, internal shocks
by faster jet blobs following slower ones, may catch up and produce internal shocks
close ($\sim 100 \rs  $) to the central BH \citep{katetal01}, which was seen in some simulations
of accretion-ejection system
\citep{lckhr16}. However, it is not plausible to form shocks closer to the jet base by this process,
because the leading blob and the one following, both
posses different but relativistic speeds, so a significant distance has to be traversed before they collide.
Is it at all possible to form shocks in the jet much closer to the central object?

As jets are supposed to originate close to the central object and from the accreting matter,
therefore the jet base should be subsonic. 
Since the jets are observed when they are actually far from the central object
and traveling at a very high speed, therefore,
one can definitely say that these jets are transonic (transits from subsonic to supersonic) in nature.
We know conical flows are smooth monotonic functions of distance, or in other words, cross
the sonic point only once \citep{m72,bm76,cr09}. Since the base of the jet is very hot, it would
be expanding very fast, and there would be very little resistance to the outflowing jet to
force it
to deviate from its conical trajectory.
However, if the jet is flowing through intense radiation field, then the jet would see a fraction of the
radiation field approaching it, and might get slowed down, to form multiple sonic point \citep{vkmc15}
and even shock \citep{fr85}.
It is also to be noted that,
most of the theoretical investigations on jets were conducted
in special relativistic regime, including \citet{fr85} and \citet{vkmc15}. The reason being that, the distances
from the central BH at which the jets are observed, the effect of gravity is negligible but the bulk speed
is relativistic. And in order to limit the forward expansion of the jet, a Newtonian form of gravitational
potential was added adhoc in the special relativistic equations of motion \citep{fr85,f96,c05,vkmc15}.
Newtonian (or, pseudo-Newtonian) gravitational potential is incompatible with special relativity which
can be argued from the equivalence principle itself. But, if we choose not to be too fussy, even
then, gluing special relativity and gravitational potentials destroy the constancy of Bernoulli parameter
in absence of dissipation, in other words, we compromise one of the constants of motion.
Moreover, \citet{fr85} obtained spiral type sonic points, and we know that spiral type sonic points
are obtained in presence of dissipation. Is this the direct fall out of combining Newtonian potential
with special relativity?
Therefore, we chose Schwarzschild metric, in order to consider gravity properly. But in keeping with most of the investigations of jet \citep{f87b,mfb02,falc96}, no particular accretion disc model is used to launch the jet.
Furthermore, radial outflows do not show any multiplicity of sonic points, then what
should generate multiple sonic points in jets. Does accretion disc shape the jet in such a fashion?  
In addition, most of the attempts on obtaining jet solutions were conducted in the
domain of fixed adiabatic index ($\Gamma$) equation of state for
jet fluid. How would a relativistic and therefore variable $\Gamma$ equation of state of the gas affect the solution? What would be the possible effect of composition in
the jet. In this paper, we would address these issues in details. 

In 
section \ref{sec2} we present simplifying assumptions, governing equations.
In section \ref{sec3}, we outline the process of generating solutions 
along with detailed discussion on nature of sonic points and shock 
conditions. Then we present results in section \ref{sec4} and finally conclude the 
analysis in section \ref{sec5}.

\section{Assumptions, governing equations and jet geometry}
\label{sec2}
Since the present study is aimed at studying jet starting very close to the central object, general relativity
is invoked.
We choose the simplest metric, \ie  Schwarzschild metric, which describes 
curved space-time around a non-rotating BH, and is given by
\be		
ds^2=-\left(1-\frac{2G\mbh}{c^2r} \right)c^2dt^2+
\left(1-\frac{2G\mbh}{c^2r}\right)^{-1}dr^2
+r^2d{\theta}^2+r^2\sin^2{\theta}d\phi^2,
\label{metric.eq}
\ee
where $r$, $\theta$ and $\phi$ are usual spherical 
coordinates,
$t$ is time and $\mbh$ is the mass of the central 
black hole. 
In the rest of the paper, we use geometric units where $G=M_B=c=1$, 
so that the units of
length and time are $\rg=GM_B/c^2$ and $\tg=GM_B/c^3$, respectively.
In this system of units, the Schwarzschild radius,
or the radius of the event horizon is $\rs=2$. Although in the rest of the paper,
we express the equations
in the geometric units (until specified otherwise), we choose to retain
the same representation for the coordinates as in equation (\ref{metric.eq}). 
The fluid jet is considered to be in steady state ({\ie} 
$\partial/\partial t=0$). Further, as the jets are collimated, we consider an on axis
({\ie} $u^\theta=u^{\phi}=\partial/\partial \theta=0$) and 
axis-symmetric ($\partial/\partial \phi=0$) jet.
Effects of radiation and magnetic field as dynamic components are being ignored for simplicity.
If the jet is very hot at the base, then radiation driving near the base will be ineffective \citep{vkmc15}. In
powerful jets, magnetic fields are likely to be aligned with the local velocity vector
and hence the magnetic force term will not arise \citep[similar to coronal holes, see][]{kh76}.
Therefore, up to a certain level of accuracy, and in order to simplify our treatment,
we ignore radiation driving and magnetic fields in the present paper and effects of these will be dealt
elsewhere.
In the advective disc model, the inner funnel or post-shock disc
(PSD) acts as the base of the jet \citep{cd07,kc13,kcm14,kc14,dcnm14,ck16,lckhr16}
and also the Comptonizing corona \citep{ct95}. 
The shape of the PSD is torus \citep[see simulations][]{dcnm14,lckhr16} and its dimension is about $\gsim$few$\times10 \rg$.
Therefore, launching the jet inside the torus shaped funnel, simultaneously
satisfies the observational requirement that the corona and the base of the jet be compact.
Having said so, we must point out that, we actually do not obtain the jet input parameters ($E$ \& ${\dot M}$) from advective accretion disc solutions, but the input parameters are supplied. This implies, any accretion disc model with compact torus like hot corona will satisfy the underlying disc model. However, advective disc model with PSD gets a special mention because possibility of hot electron distribution close to the
central object, is inbuilt to the model. The only role of the disc considered here, is to confine the jet flow boundary at the base, for one of the jet model (M2) considered in this paper.
Since the exact method of
how the jet originates from the disc is not being considered in this paper, so the jet input parameters are actually free parameters independent of the disc solutions. This paper is an exploratory study of the role of jet flow geometry close to the base,
on jet solutions and therefore we present all possible jet solutions. 

Observations show that the core temperatures of powerful AGN jets are estimated to be quite high
\citep{mo96}. So the jets are hot to start with in this paper too. The advective disc model, as in most disc models, do come with a variety of inner disc temperatures. Simulations of advective discs
for high viscosity parameter produced $T \gsim 10^{12}$K in the PSD \citep{lckhr16}. Moreover in presence
of viscous dissipation in curved space-time, the Bernoulli parameter ($-hu_t$) may increase by more than
$20\%$ of its value
at large distance and produce very high temperatures in the PSD
\citep{ck16}.
For highly rotating BHs too, the temperatures of the inner disc easily approaches $10^{12}$ K.
It must also be remembered that inner regions of the accretion disc can be heated by Ohmic dissipation,
reconnection, turbulence heating or MHD wave dissipation may heat up the inner disc or the base of the jet \citep{be03}. High temperatures in the accretion disc can
induce exothermic nucleosynthesis too \citep{cja87,hp08}. All these processes taken together in an advective disc
will produce very hot jet base. 
We do not specify the exact processes that will produce very hot jet base, but would like to emphasize
that it is quite possible to achieve so.  
One may also wonder, that if the jets are indeed launched from the disc, how justified is it to consider
non-rotating jets. Phenomenologically speaking, if jets have a lot of rotation then it would not flow around the axis of symmetry and therefore, either it has to be launched with less angular momentum or, has to loose most of the angular momentum with which it is launched. 
It has been shown that viscous transport removes significant angular momentum
of the collimated outflow close to the axis \citep{lckhr16}.  
Since the jet is launched with low angular momentum and it is further removed by viscosity or by the presence of
magnetic field, therefore, the assumption of non-rotating hot jet is quite feasible. Incidentally, similar to
this study, there are many theoretical studies of jets which have been undertaken
under similar assumptions of non-rotating, hot jets at the base \citep{f87b,mfb02,falc96}.

\begin{figure}
\begin{center}
\includegraphics[width=12.cm]{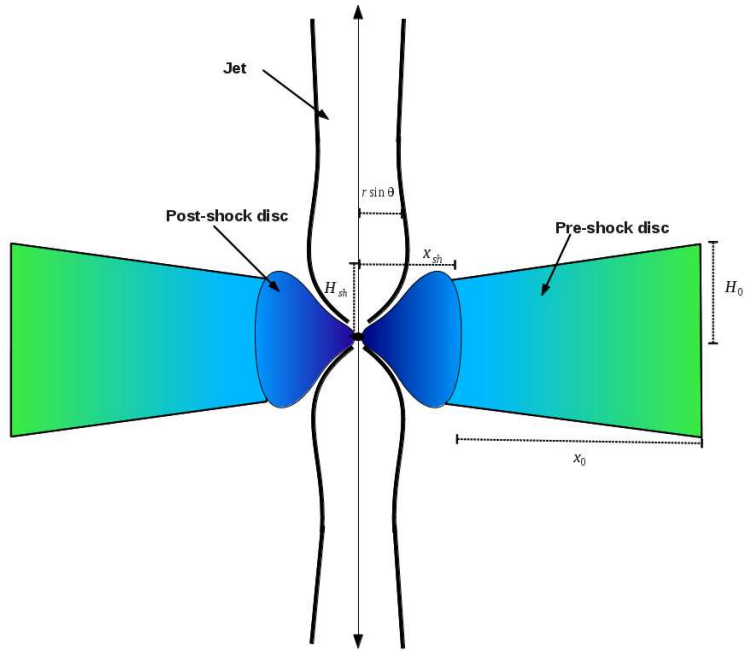}
\vskip -0.5cm
 \caption{Cartoon diagram of accretion disc-jet geometry of model M2.
The pre-shock and the 
post-shock disc are shown. 
On the figure we mark the shock location $\xsh$, half-height of the 
shock $H_{\rm sh}$ , the outer edge of the
disc $x_{\rm o}$ and half height at $x_{\rm o}$ is $H_{\rm o}$.}
\label{lab:fig1}
 \end{center}
\end{figure}

\subsection{Geometry of the jet}
\label{sec:jet_geom}
Observations of extra-galactic jets show high degree of collimation, so it is quite
common in jet models to consider conical jets with small opening angle. We considered two models of jets, the first model being a jet
with conical cross-section and we call this model as M1.
However, the jet at its base is very hot and subsonic, and since 
the pressure gradient is isotropic, the jet will expand in all directions.
The walls of the inner funnel of the PSD will provide natural collimation
of the jet flow near its base. If the base of the jet is very energetic, then it is quite likely
to become transonic within the funnel of the PSD (see Fig. \ref{lab:fig1}).
In a magneto-hydrodynamic simulations
of jets, \citet{kms02} showed that the jet indeed flows through the open field line region, but at a certain distance above the torus shaped disc,
the jet surface is pushed towards the axis and beyond that the jet expands again, resulting into a converging-diverging cross-section. 
The cross-section adopted for Model M2 in this paper,
is inspired by these kinds of jet simulations. We ensured that the variation of the geometry of M2 is smooth and slowly varying, such that the solutions are
not crucially dependent on the particular shape of the flow geometry.

The general form of jet cross section or, ${\cal A}(r)$ is
\be
{\cal A}={\cal G}r^2,~~ \mbox{where,~}{\cal G}=2\pi(1-{\rm cos}\theta);~~\theta \mbox{ is polar angle at }r
\label{jetgeom.eq}
\ee 
For the first model M1: 
\be
cos\theta=\frac{-(C m)/r+{\sqrt{-C^2/r^2+m^2+1}}}{(m^2+1)},
\label{con2a.eq}
\ee
where, $C$ and $m(= {\tan {\theta_0}})$ are the constant intercept and slope of the jet boundary with the equatorial plane, with $\theta_0$
being the constant opening angle with the axis of the jet. The value of the constant
intercept might be
$C=0$ or $C \neq 0$, either way, the solutions are qualitatively same and we take $C=0$.

The second model M2 mimics a geometry whose outer boundary is the funnel shaped surface
of the accretion disc at the jet base i. e., $d{\cal A}/dr >0$
(see, Fig. \ref{lab:accretion_disc_jet_plot}a). As the jet
leaves the funnel shaped region of PSD, the rate of increase of the cross-section gets reduced
$d{\cal A}/dr \sim 0$. It again expands
and finally becomes conical at large distances, where $d{\cal A}/dr \propto r$.
The functional form of $r$ and $\theta$ of the jet geometry for the model M2 is given by,
\be
r~{\rm sin}\theta=\frac{k_1d_1(r-n)}{1+d_2(r-n)^2}+m_{\infty}(r-n)+k_2,
\label{geomvar.eq}
\ee
where, $k_1=5\xsh/\pi$, $k_2=\xsh/2$, $d_1=0.05$, $d_2=0.0104$, $m_{\infty}=0.2$ and $n=5$.
A schematic diagram of the geometry of the M2 jet model and disc is shown in Fig. \ref{lab:fig1}, where $\xsh$ is the shock
in accretion, or in other words, the length scale of the inner torus like region.
In Eq. (\ref{geomvar.eq}), $k_1$, $k_2$
are parameters which influence the shape of jet geometry at the base, while $d_1$, $d_2,~m_{\infty}$ and $n$
are constants, which together with $k_1$ and $k_2$, shape the jet geometry. Here, we assume at large distances
the jet is conical and $m_\infty(= {\tan {\theta_{\infty}}})$ is the gradient which corresponds to the terminal
opening angle to be 11$^\circ$.
The size of the PSD or $\xsh$ influences the jet geometry and
the jet geometry at the base is shaped by the shape of the PSD as shown in appendix (\ref{disc_structure}).
A typical jet geometry for a given set of accretion solution is plotted in Fig. \ref{lab:accretion_disc_jet_plot}.

\subsection {Equations of motion of the jet}

\subsubsection {Equation of state}
\label{sbsbsec2.1.1}
Hydrodynamic equations of motion are solved with the help of a
closure relation between internal energy density, pressure and mass
density ($e,~p,~\rho$, respectively) of the fluid,
called the equation of state (EoS).
An EoS for multispecies, relativistic
flow proposed by \citet{c08,cr09} is adopted, and is 
given by,
\begin{equation}
e=n_{e^-}m_ec^2f,
\label{eos.eq}
\end{equation}
were $n_{e^{-}}$ is the electron number density and $f$ is given by
\begin{equation}
f=(2-\xi)\left[1+\Theta\left(\frac{9\Theta+3}{3\Theta+2}\right)\right]
+\xi\left[\frac{1}{\eta}+\Theta\left(\frac{9\Theta+3/\eta}{3\Theta+2/\eta}
\right)\right].
\label{eos2.eq}
\end{equation}
Here, non-dimensional temperature is defined as
$\Theta=kT/(m_ec^2)$, $k$ is the Boltzmann constant and
$\xi = n_{p^{+}}/n_{e^{-}}$ is the relative proportion of 
protons with respect to the number density of electrons.
The mass ratio of electron and proton is 
$\eta = m_{e}/ m_{p^{+}}$. It is easy to see that
by putting $\xi=0$, we generate EoS for relativistic 
$\eel$ plasma \citep{rcc06}.
The expressions of the polytropic index $N$, adiabatic 
index $\Gamma$ and
adiabatic sound speed $a$ are given by
\begin{equation}
N=\frac{1}{2}\frac{df}{d\Theta} ;~~ \Gamma=1+\frac{1}{N} ; ~~
a^2=\frac{\Gamma p}{e+p}=\frac{2 \Gamma \Theta}
{f+2\Theta}.
\label{sound.eq}
\end{equation}
This EoS is an approximated one, and the comparison 
with the exact one shows that this EoS
is very accurate \citep[Appendix C of][]{vkmc15}. 
Additionally, being algebraic and avoiding the presence
of complicated special functions,
this EoS is very easy to be
implemented in simulation codes, as well as, be used 
in analytic investigations \citep{cr09,cc11,rcc06,crj13}.

\subsubsection{Equations of motion}
\label{sec:eom}
The energy momentum tensor of the jet matter is given by
\begin{equation}
T^{\alpha \beta}=(e+p)u^{\alpha}u^{\beta}+pg^{\alpha \beta}
\end{equation}
where the metric
tensor components are given by $g^{\alpha \beta}$ 
and $u^{\alpha}$ represents	 four velocity.

The equations of motion are given by
\begin{equation}
T^{\alpha \beta}_{;\beta}=0~~~~  \mbox{and} ~~~~
(\rho u^{\beta})_{; \beta}=0,
\label{eqnmot.eq}
\end{equation}
The first of which is energy-momentum conservation 
equation and second is continuity equation. From the 
above equation, the $i^{th}$ component of the momentum conservation equation
is obtained by operating the projection 
tensor on the first of equation (\ref{eqnmot.eq}),
{\ie}
\begin{equation}
(g^{i}_{\alpha}+u^iu_\alpha)T^{\alpha \beta}_{{;\beta}}=0
\label{genmomb.eq}
\end{equation}
Similarly, the energy conservation equation is 
obtained by taking 
\begin{equation}
u_{\alpha}T^{\alpha \beta}_{{;\beta}}=0
\label{genfstlaw.eq}
\end{equation}
For an on-axis jet, equations (\ref{genmomb.eq}, \ref{genfstlaw.eq})
becomes; 
\begin{equation}
u^r\frac{du^r}{dr}+\frac{1}{r^2}=-\left(1-\frac{2}{r}+u^ru^r\right)
\frac{1}{e+p}\frac{dp}{dr},
\label{eu1con.eq}
\end{equation}
\begin{equation}
\frac{de}{dr}-\frac{e+p} {\rho}\frac{d\rho}{dr}=0,
\label{en1con.eq}
\end{equation}
While the second of equation (\ref{eqnmot.eq}) when integrated becomes
mass outflow rate equation,
\begin {equation}
\dot {M}_{\rm {out}}=\rho u^r {\cal A}.
\label{mdotout.eq}
\end {equation}
Here, ${\cal A}$ is the cross-section area of the jet. 
The differential form of the outflow rate equation is, 
\begin{equation}
\frac{1}{{\rho}}\frac{d{\rho}}{dr}=-\frac{1}{{\cal A}}\frac{d{\cal A}}{dr}
-\frac{1}{u^r}\frac{du^r}{dr}.
\label{con1con.eq}
\end{equation}
By using equation (\ref{mdotout.eq}), pressure $p$ can
be given as
\be
p=\frac{2\Theta \rho}{\tau}=\frac{2\Theta \dot {M}_{\rm {out}}}{\tau u^r {\cal A}}
\label{pressure.eq}
\ee
Here ${\cal A}$ is the cross section of the jet 
(section \ref{sec:jet_geom}) and $\tau=(2-\xi+\xi/\eta)$.
Equations (\ref{eu1con.eq}-\ref{en1con.eq}), with the help of equation (\ref{con1con.eq}), 
are simplified to
\begin{equation}
\gamma^2v\left(1-\frac{a^2}{v^2}\right)\frac{dv}
{dr}=\left[a^2\left\{\frac{1}{r(r-2)}+\frac{1}{{\cal A}}
\frac{d{\cal A}}{dr}\right\}-\frac{1}{r(r-2)}\right]
\label{dvdr.eq}
\end{equation}
and
\begin{equation}
\frac{d{\Theta}}{dr}=-\frac{{\Theta}}{N}\left[ \frac{{\gamma}
^2}{v}\left(\frac{dv}{dr}\right)+\frac{1}{r(r-2)}
+\frac{1}{{\cal A}}\frac{d{\cal A}}{dr}\right]
\label{dthdr.eq}
\end{equation}
Here the three-velocity $v$ is given by
$v^2=-u_iu^i/u_tu^t=-u_ru^r/u_tu^t$, {\ie}
$u^r=\sqrt{g^{rr}}{\gamma}v$ and $\gamma^2=-u_tu^t$ is the Lorentz factor.
All the information like jet speed, temperature, sound speed, adiabatic index, 
polytropic index as functions of spatial distance, can be obtained by integrating equations (\ref{dvdr.eq}-\ref{dthdr.eq}).

A comparison with De Laval Nozzle helps us to understand the nature of the equations
and the expected nature of the solutions.
The non-relativistic version of De Laval Nozzle (DLN) is obtained by considering $v<< 1$ and
$r \rightarrow$large in equation (\ref{dvdr.eq}), which
means the first and the third term in r. h. s drop off and $\gamma~\rightarrow~1$, but for the
special relativistic (SR) version there is no constrain on $\gamma$.
In the subsonic regime ($v<a$), the jet will accelerate if $d{\cal A}/dr<0$ (converging cross-section), both in
the non-relativistic and in SR version of DLN problem. While in the supersonic
regime ($v>a$) the jet accelerate if $d{\cal A}/dr>0$ (expanding cross-section), and forms
the sonic point where $d{\cal A}/dr=0$.
Therefore, a pinch in the flow
geometry (i. e., a convergent-divergent cross section)
makes a subsonic flow to become transonic.
However, in presence of gravity, the first and the third terms cannot be ignored.
The third term is the purely gravity term, while the first term is the coupling between gravity and the thermal term.
Therefore, the gravity and the flow's thermal energy
now compete with the cross-section term to influence the jet solution.
The third term is negative, the first term is positive definite, the middle term
may either be positive or negative. However, near the horizon the third dominates the other two
and even for a conical flow ($d{\cal A}/dr>0$), the r. h. s of equation (\ref{dvdr.eq}) will be negative. Hence a subsonic jet will accelerate up to the sonic point $r_c$, where r. h. s becomes
zero. For $r>r_c$ the flow is supersonic and still the jet accelerates because at those distances
the r. h. s of equation (\ref{dvdr.eq}) is positive. Therefore, unlike the typical DLN problem,
in presence of gravity, it is not mandatory that $d{\cal A}/dr=0$ so that an $r_c$ may form,
i. e., gravity ensures the formation of the sonic point. But if the magnitude of
$d{\cal A}/dr$ changes drastically, even without changing sign, the interplay with the
gravity term may ensure formation of multiple sonic points. 

A physical system becomes tractable when solutions are described in terms of their constants of motion.
Using a time like Killing vector
$\zeta_\nu=(1,0,0,0)$, the equation of motion (equation \ref{eqnmot.eq}) becomes,
\begin{eqnarray*}
\left(T^{\mu \nu}\zeta_{\nu}\right)_{;\mu}=J^\mu_{;\mu}=0; \\
{\rm or}, ~\left(\sqrt{-g}J^r\right)_{,r}=\left(r^2{\rm sin}\theta T^r_t\right)_{,r}=0.
\end{eqnarray*}
Integrating the above equation,
we obtain the negative of the energy flux as a constant of motion,
\be
{\cal{A}} (e+p)u^ru_t=-{\dot E}={\rm constant}
\label{energflux.eq}
\ee
The relativistic Bernoulli equation or, the specific energy of the jet,
is obtained by dividing equation (\ref{energflux.eq}) by equation(\ref{mdotout.eq})
\begin{equation}
E=\frac{{\dot E}}{{\dot M}_{\rm out}}=-hu_t. 
\label{enr.eq}
\end{equation}
Here, $h =({e+p})/{\rho}=({f+2\Theta})/{\tau}$ is the specific enthalpy
of the fluid and $u_t=-{\gamma}\sqrt{(1-2/r)}$.
The kinetic power of the jet is the energy flux through the cross-section
\be
\Lj={\dot E}={\dot M}_{\rm out}E
\label{ljet.eq}
\ee
If only the entropy equation (\ref{en1con.eq}) is integrated, then we obtain the adiabatic relation
which is equivalent to
$p\propto \rho^{\Gamma}$ for constant $\Gamma$ flow \citep{kscc13},
\be
\rho={\cal C}\mbox{exp}({\rm k}_3) \Theta^{3/2}(3\Theta+2)^{{\rm k}_1}
(3\Theta+2/\eta)^{{\rm k}_2},
\label{rho.eq}
\ee
where, ${\rm k}_1=3(2-\xi)/4$, ${\rm k}_2=3\xi/4$, ${\rm k}_3=(f-\tau)/(2\Theta)$ and ${\cal C}$ is the constant
of entropy.
If we substitute $\rho$ from the above equation into 
equation (\ref{mdotout.eq}), we get the expression 
for entropy-outflow
rate \citep{kscc13},
\begin{equation}
{\dot {\cal M}}=\frac{{\dot M}_{\rm out}}{{\rm geom. const.}
{\cal C}}=\mbox{exp}({\rm k}_3) \Theta^{3/2}(3\Theta+2)
^{{\rm k}_1}
(3\Theta+2/\eta)^{{\rm k}_2}u^r{\cal A}
\label{entacc.eq}
\end{equation}
Equations (\ref{entacc.eq}) and (\ref{enr.eq}) are 
measures of entropy and energy of the flow that remain 
constant along a streamline. However, at the shock, 
there is a discontinuous jump of ${\dot {\cal M}}$.
\section{Solution method}
\label{sec3}

\begin{figure}
\begin{center}
 \includegraphics[trim={0 0 0 2.4cm},clip,width=12cm]{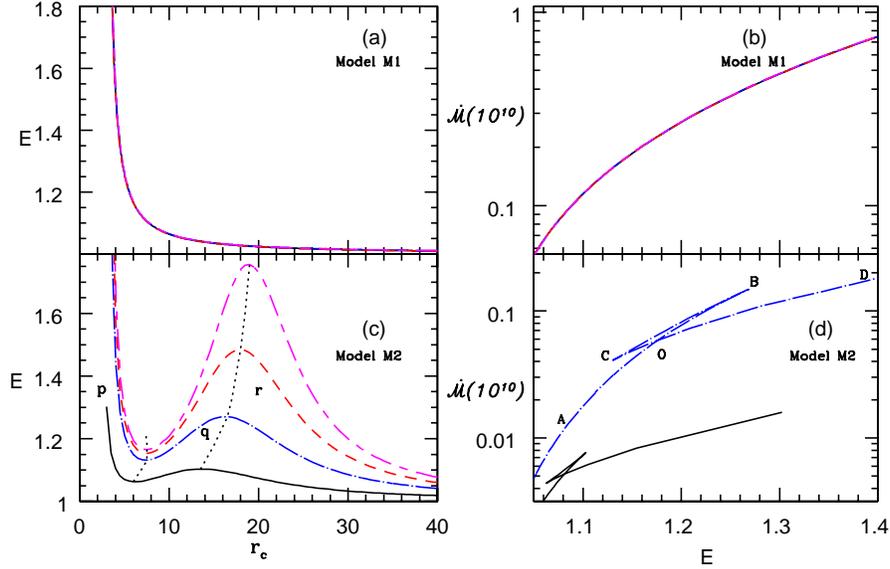}
 \caption{Variation of $E$ with $r_c$ (a, c) and ${\md}$ with $E$ (b, d) for the two models M1 (a, b)
 and M2 (c, d). Each curve represent $x_{\rm sh}=12$ (long-short dash, magenta), $10$ 
 (dash, red), $7$ (dash-dot, blue),
 $3$ (solid, black). (d) $\md~(r_c)$-$E~(r_c)$ are plotted
 for only two $\xsh=3$ (solid, black) and $\xsh=7$ (dash-dot, blue). The composition of the jet is for $\xi=1.0$.
 }
\label{lab:fig2}
 \end{center}
\end{figure}
\subsection{Sonic point conditions}
\label{sbsec3.1}
Jets originate from a region in the 
accretion disc, which is close
to the central object, where, the jet is subsonic and very hot.
The thermal gradient being very strong, works against the gravity and
power the jet to higher velocities and in the process, $v$ crosses
the local sound speed $a$ at the sonic point $r_c$, which makes the jet supersonic.
The sonic point is also a critical point because, at $r_c$
equation (\ref{dvdr.eq}) 
takes the form $dv/dr\rightarrow 0/0$.
This gives us the sonic point conditions,
\be 
 v_c=a_c
\ee 
\begin{equation}
a_c^2=\left[1+r_c(r_c-2)\left(\frac{1}{{\cal A}}\frac{d{\cal A}}{dr}\right)_c\right]^{-1}\
\label{sonic.eq}
\end{equation}
The $dv/dr|_c$ is calculated by employing the L'Hospital's 
rule at $r_c$ and solving the resulting
quadratic equation for $dv/dr|_c$. The quadratic 
equation can admit two complex roots leading to
the either $O$ type (or `centre' type) or `spiral' type
sonic points, or two real roots but with opposite signs
(called $X$ or `saddle' type sonic points), or real roots 
with same sign (known as nodal type sonic point).\\
So for a given set of flow variables at the jet base, a 
unique solution will pass through
the sonic point(s) determined by the 
entropy ${\dot {\cal M}}$ and energy $E$ of the flow.
Model M1 is independent of the shock location $\xsh$ in the accretion disc,
but model M2 depends on $\xsh$.
In Figs. (\ref{lab:fig2}a, b), we plot the sonic point properties of
the jet for model M1, and in Figs. (\ref{lab:fig2}c, d) we plot the
sonic point properties of M2. Each curve is plotted for 
$x_{\rm sh}=12$ (long-short dash, magenta), $10$ 
(dash, red), $7$ (dash-dot, blue), $3$ (solid, black).
The Bernoulli parameter of the jet $E$ is plotted as a function of $r_c$ in Fig. (\ref{lab:fig2}a, c).
In Fig. (\ref{lab:fig2}b, d), ${\dot {\cal M}}$ is plotted as a function of $E$ at the sonic points.
The jet model M1 is a conical flow and therefore, is independent of the
accretion disc geometry. So,
curves corresponding to various values of $\xsh$ coincide with each other in Figs. \ref{lab:fig2}a and b.
Moreover, both $E$ and ${\dot {\cal M}}$ are monotonic functions of $r_c$. In other words, a flow with a given $E$ will have one sonic point, and the transonic solution will correspond to
one value of entropy, or ${\dot {\cal M}}$.
The situation is different for model M2. As $\xsh$ is increased from $3,~7,~10,~12$, the $E$ versus $r_c$ plot
increasingly deviates from monotonicity and produces multiple sonic points
in larger range of $E$ (Fig. \ref{lab:fig2}c). 
For small values of $\xsh$ the jet cross-section is very close to the conical geometry and therefore, multiplicity of sonic points is obtained in a limited range of $E$.
It must be noted that, for a given $\xsh$, jets with very high and low values of $E$
form single sonic points. The range of $E$ within which multiple sonic points may form,
increases with increasing $\xsh$. The reason M1 has only one $r_c$, is amply clear from
our discussion on DLN in section \ref{sec:eom}.
We know that, if r. h. s of equation (\ref{dvdr.eq}) is zero, a sonic point is formed.
Since ${\cal A}^{-1}d{\cal A}/dr=2/r$ is always positive, the r. h. s 
becomes zero only due to gravity, and therefore, there is only one $r_c$ for M1.

For M2, the cross-section near the base expands faster than a conical cross-section,
therefore, the first two terms in r.h.s of equation (\ref{dvdr.eq}) competes with gravity.
As a result, the jet rapidly accelerates to cross the sonic point withing the funnel like
region of PSD.
But as the jet crosses the height of PSD, the expansion is arrested and at some height
${\cal A}^{-1}d{\cal A}/dr \sim 0$. If this happens closer to the jet base
then the gravity will again make the r. h. s of the equation (\ref{dvdr.eq}) zero,
causing the formation of multiple sonic points. 
For low values of $E$ in M2, the thermal driving
is weak and so the sonic point forms at large distances. At those distances
${\cal A}$ becomes almost conical and therefore, for reasons cited above, 
the jet has only one sonic point.
If $E$ is very high, then the strong thermal driving makes the jet transonic
at a distance very close to the jet base.
For such flow, the thermal driving remains strong enough even in the supersonic domain, which
negates the effect of changing $d{\cal A}/dr$ and do not produce more sonic points.
For intermediate values of $E$, the jet becomes transonic at slightly larger distances.
For these flows,
the thermal
driving in the supersonic region becomes weaker and at the same time, the expansion of the jet cross section
term decreases i. e.,  
${\cal A}^{-1}d{\cal A}/dr \sim 0$. At those distances, the gravity again becomes dominant
than the other two terms, which reduce the
r. h. s of equation(\ref{dvdr.eq}) and makes it zero to produce multiple sonic points.
In Fig. (\ref{lab:fig2}c), the maxima and minima of $E$ is the 
range which admits multiple sonic points.
We plotted the locus of the maxima and minima with a dotted line, and then divided the region
as `p', `q' and `r'. Region `p' harbours inner X-type sonic point, region `q' harbours O-type
sonic point and region `r' harbours outer X-type sonic points.
Figure (\ref{lab:fig2}d) 
is the knot diagram \citep[similar to `kite-tail' for accretion, see][]{kscc13} between $E$ and 
${\dot {\cal M}}$ evaluated at $r_c$, for two values of $x_{\rm sh}=3$ (solid, black) and $\xsh=7$ (dashed-dot, blue).
For $\xsh=7$ (dashed-dot, blue),
the top flat line of the knot (BC) represents the O-type sonic points from region `q'
of Fig (\ref{lab:fig2}d). Similarly,
AB represents outer X-type sonic points from the region `r'.
And CD gives the values of $E$ and ${\dot {\cal M}}$ for which only
inner X-type sonic points (region `p') exists. If the coordinates of the turning/end points of the curve
be marked as ${\md}_{\rm B},~E_{\rm B}$ and so on, then it is clear that for $\xsh=7$, multiple sonic points
form for jet parameter $E_{\rm C}\leq E \leq E_{\rm B}$. At the crossing point `O', the entropy of both the
sonic points are same.
The plot for $\xsh=3$ (solid, black) is plotted as a comparison, and shows that if the shock in accretion is formed
close to the central object, then multiple sonic points in jets are formed for moderate range of $E$, a fact also
quite clear from Fig. (\ref{lab:fig2}c).

\subsection{Shock conditions}
One of the major outcomes of existence of multiple 
sonic points in the jet, is the possibility of 
formation of shocks in the flow. At the shock,
the flow makes a discontinuous jump in density, pressure and velocity.
The relativistic Rankine-Hugoniot conditions relate the flow quantities across the
shock jump and they are \citep{t48,cc11}
\begin{equation}
  [{\rho}u^r]=0,
  \label{sk1.eq}
\end{equation}
\begin{equation}
   [T^{tr}]=[(e+p)u^tu^r]=0,
   \label{sk2.eq}
\end{equation}
and
\begin{equation}
[T^{rr}]=[(e+p)u^ru^r+pg^{rr}]=0
\label{sk3.eq}
\end{equation}
The square brackets denote the difference 
of quantities across the shock, i.e. 
$[Q]=Q_2-Q_1$ 
with $Q_2$ and  $Q_1$ being 
the quantities after and before the shock respectively.\\
Dividing equation (\ref{sk2.eq}) by equation 
(\ref{sk1.eq}) and then simplifying, we obtain
\be
[E]=[hu_t]=0
\label{sk4.eq}
\ee
It merely states that the energy remains 
conserved across the shock. Further, dividing 
(\ref{sk3.eq}) by (\ref{sk1.eq}) and a little 
algebra leads to
\be
\left[\sqrt{g^{rr}}\left(h \gamma v+\frac{2 \Theta}{\tau \gamma v}\right)\right]=0
\label{sk5.eq}
\ee
We check for shock conditions (equations \ref{sk4.eq}, \ref{sk5.eq})
as we solve the equations of motion
of the jet.
The strength of the shock is measured by two parameters  
compression ratio ($R$) and shock strength ($S$). 
$R$ and $S$ are ratios of densities and Mach numbers ($M$) across
the shock (at $r=r_{\rm sh}$). In relativistic case, according to equation
 (\ref{sk1.eq}) $R$ is obtained as,
\be 
R=\frac{\rho_+}{\rho_-}=\frac{u_-^r}{u_+^r}=\frac{\gamma_- v_-}{\gamma_+ v_+},
\label{compress.eq}
\ee
where, $+$ and $-$ stands for quantities at post-shock and pre-shock flows, respectively. 
Similarly, $S$ is defined as,
\be
S=\frac{M_-}{M_+}=\frac{v_-a_+}{v_+a_-}
\label{shokstrnth.eq}
\ee

\section{Results}
\label{results}
\label{sec4}
In this paper, we study relativistic jets flowing through two types of geometries:
(i) model M1: conical jets and (ii) model M2: jets through a variable and non-conical cross-section,
as described in section \ref{sec:jet_geom}.

\subsection{Model M1 : Conical jets}
\label{sec:m1}
\begin{figure}
\begin{center}
 \includegraphics[trim={0.0 6.0cm 0 0},clip,width=10cm]{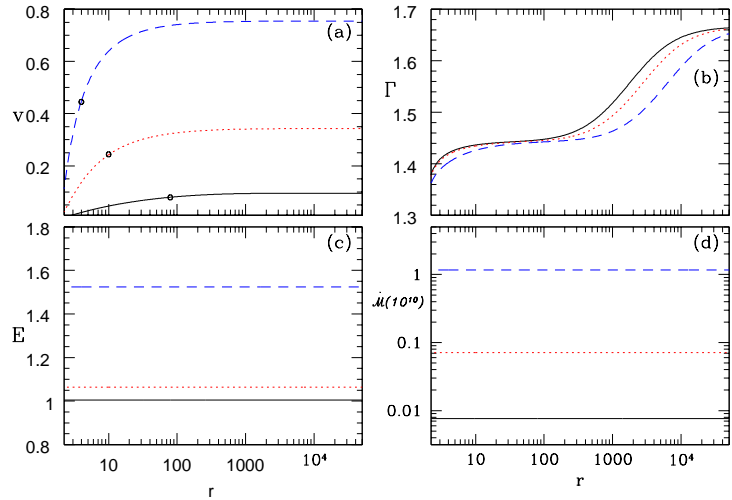}
 \hskip -5.0cm
 \caption{Model M1: Variation of (a) three velocity $v$, (b) 
adiabatic index $\Gamma$, (c) energy parameter $E$ and (d) entropy $\md$ as functions of $r$.
 The shock in accretion $x_{\rm sh}=12$ and flow composition $\xi=1.0$. 
 Each curve is characterized for $E=1.0045$ (solid, black), 
 $E=1.064$ (dotted, red) and $E=1.525$ (dashed, blue). 
 Black open circles in (a) show the location of sonic points.}
\label{lab:fig3}
 \end{center}
\end{figure}
Inducting equation (\ref{con2a.eq}) into (\ref{jetgeom.eq}), 
we get a spherically outflowing jet, which, for $C=0$ gives a constant 
$\theta(=\theta_0=11^\circ)$. The jet geometry is such that ${\cal A} \propto r^2$. 
Keeping $\xi=1$, each curve for model M1 is plotted 
for $E=1.0045$ (solid black), $E=1.064$ (dotted red) 
and $E=1.525$ (long-dashed blue) and are shown in Figs. (\ref{lab:fig3}a-d). 
Open circles denote the location of sonic points. For higher values of $E$, the jet terminal speed is also 
higher (Fig. \ref{lab:fig3}a). Higher $E$ also produce hotter flow. So at any given $r$, $\Gamma$ is lesser for higher E (Fig. \ref{lab:fig3}b)).
Since the jet is smooth and adiabatic, so $E$ (Fig. \ref{lab:fig3}c) and $\md$ (Fig. \ref{lab:fig3}d)
remain constant.
The variable nature of $\Gamma$ is clearly shown in Fig. (\ref{lab:fig3}b),
which starts from a value slightly above $4/3$ (hot base) and at large distance it approaches 
$5/3$, as the jet gets colder. 
As discussed in sections \ref{sec:eom} and \ref{sbsec3.1}, for all possible 
parameters, this geometry gives smooth solutions with
only single sonic point until and unless M1 jet interacts with the ambient medium.

\subsection{Model M2}
\label{sec:m2}
\begin{figure}
\begin{center}
 \includegraphics[width=14cm]{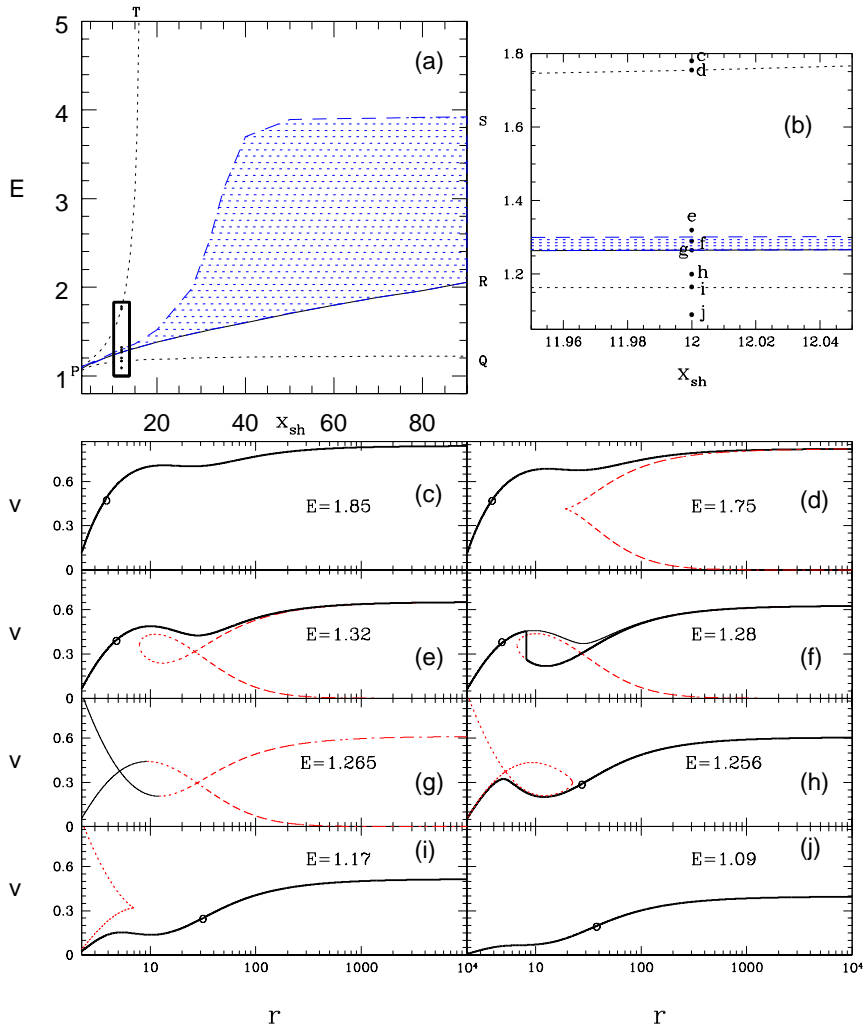}
 \caption{(a) $E$---$\xsh$ parameter space for $\xi=1$. PQRSTP is the region of multiple sonic point.
PR is the locus of points on which the two sonic points have same entropy. Shaded region PRSP permits stable shock transition. (b) Zoomed
parameter space, which shows positions `c'-`j' on $\xsh=12$ for which all typical solutions are
plotted in the panels (c)-(j). Circles on global solutions (solid) and as crossings in closed solutions (dashed)
are the location of sonic points.}
\label{lab:fig4}
 \end{center}
\end{figure}

In this section, we discuss all possible solutions associated with the jet model M2.
The sonic point analysis of the jet model M2 showed that for larger values of $\xsh$,
multiple sonic point may form in jets for a larger range of $E$ (Fig. \ref{lab:fig2}c). In Fig. (\ref{lab:fig4}a),
we plot the multiple sonic point region (PQRSTP) bounded by the dotted line. Dotted lines are same as those on Fig. (\ref{lab:fig2}c),
obtained by connecting the maxima and minima of $E$ versus $r_c$ plot. 
Jets with all $E$, $\xsh$ values within PQRSTP will harbour multiple sonic points, which is similar to
the bounded region of the energy-angular momentum space for accretion disc \citep[see, Fig. 4 of][]{ck16}.
The central solid line (PR) is the set of all $E,~\xsh$ which harbour three sonic points, but the entropy is same for both the 
inner and outer sonic points. In region QPRQ, the entropy of the inner sonic
points is higher than the outer sonic point and in TPRST, it is vice versa. The zoomed part of the parameter space around $\xsh=12$ is shown in (\ref{lab:fig4}b) and marked locations
from `c' to `j'. The solutions corresponding to the values of $E$ (marked in Fig.\ref{lab:fig4}b) and $\xsh=12$ are plotted in panels
Fig. (\ref{lab:fig4}c---j). For higher energies $E=1.85$ ($>1.75$ left side of PT), only one X-type sonic point (circle) is possible close to the
BH (Fig. \ref{lab:fig4}c). Due to stronger thermal driving the jet
accelerates and becomes transonic close to the
BH. For a slightly lower $E(=1.75)$, there are two sonic points, the solution (solid) through the inner one (shown by a small circle)
is a global solution, while the second solution (dashed) terminates at the outer sonic point (crossing point) is not global
(Fig. \ref{lab:fig4}d).  For lower $E~(=1.32)$, the solution through outer sonic point is $\alpha$ type
(dashed) and has higher entropy. An $\alpha$ type solution is the one which makes a closed loop
at $r < r_c$ and has one subsonic and another supersonic branches starting from $r_c$ outwards extending up to infinity (Fig. \ref{lab:fig4}d).
The jet matter starting from the base,
can only flow out through the inner sonic point (solid), but cannot jump onto the higher entropy solution because
shock conditions are not satisfied (Fig. \ref{lab:fig4}e). However, for $E=1.28$, the entropy difference between
inner and outer sonic points are exactly such that, matter through the inner sonic point jumps to the solution through outer
sonic point at the jet shock or $\rsh$ (Fig. \ref{lab:fig4}f).
Solution for $E=1.265$ is on PR and produces inner and outer sonic points with the same entropy
(Fig. \ref{lab:fig4}g).
Figures (\ref{lab:fig4}c ---g) are parameters lying in the TPRST. For flows with even
lower energy $E=1.256$, the entropy condition of the two physical
sonic points reverses. In this case, the entropy of the inner sonic point is higher than the outer one. So, although multiple sonic points exist but no shock in jet
is possible (Fig. \ref{lab:fig4}h) and the jet flows out through the outer sonic point.
In Fig. (\ref{lab:fig4}i), the energy is $E=1.17$ and the solution is almost the mirror image of Fig. (\ref{lab:fig4}d).
Figures (\ref{lab:fig4}h, i) belong to QPRQ region. For even lower energy i. e.,
$E=1.09$ a much weaker jet flows out through the
only sonic point available at a larger distance from the compact object (Fig. \ref{lab:fig4}j).
\begin{figure}
\begin{center}
 \includegraphics[width=13cm]{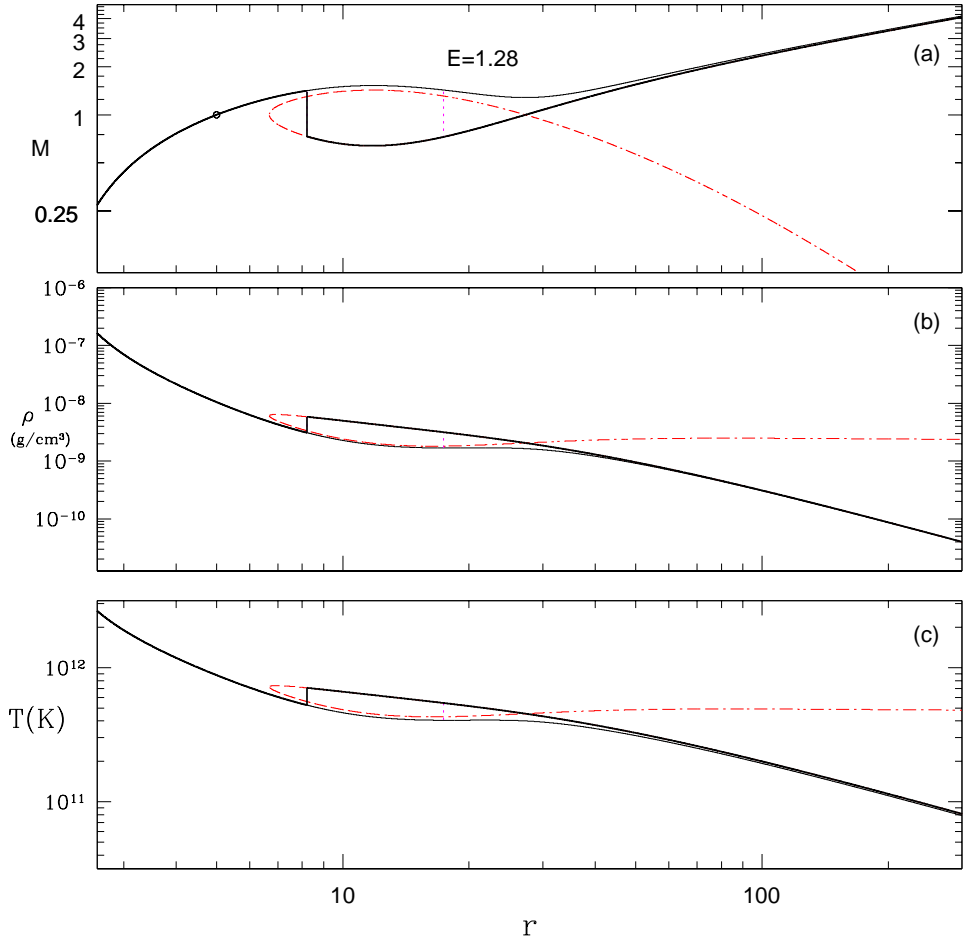}
 \caption{Variation of (a) Mach number $M$, (b) $\rho$ and (c) $T$ (in Kelvin)
 of a shocked jet solution. The density profile is plotted
 by assuming $M_{\rm BH}=10M_\odot$ and ${\dot M}=0.01{\dot M}_{\rm Edd}$. All the plots are obtained for $E=1.28$.}
\label{lab:fig5}
 \end{center}
\end{figure}

In Figs. \ref{lab:fig5}a-c, we plot the solution of the inner region of a shocked jet, where the inner boundary
values are clearly seen for a particular solution. The solid curve represents the physical solution, and the
dotted curve is a possible multi valued solution which can only be accessed in presence of a jet shock.
The base temperature and density (in physical units) are quite similar to the ones obtained in advective accretion disc solutions.

\begin{figure}
\begin{center}
 \includegraphics[width=13cm]{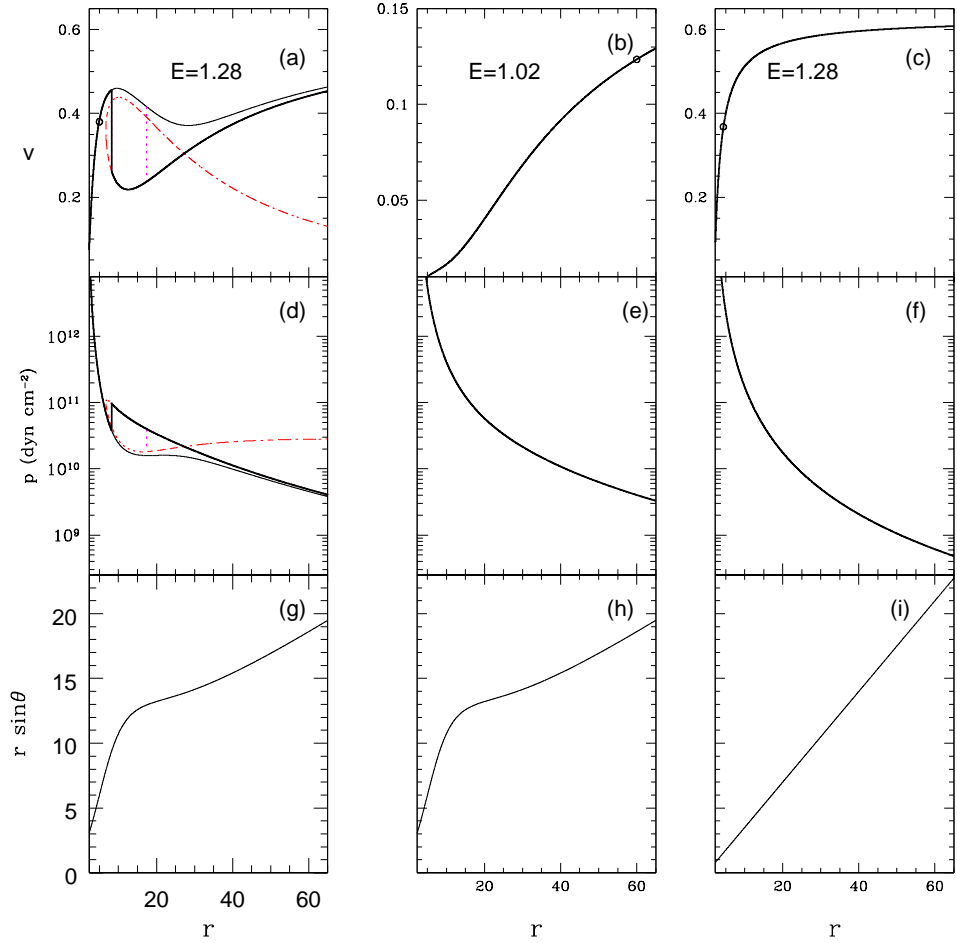}
 \caption{Three velocity $v$ (a, b, c), pressure $p$ in physical units (d, e, f)
 and jet cross section (g, h, i) are plotted
 as a function of $r$. Jet M2 are presented in the left and middle panels.
 Jet M1 is presented in the right panel. Thin-solid curve is a physical solution
jet do not follow due to shock.
Vertical lines show
stable (solid) and unstable (dotted) shocks.}
\label{lab:fig6}
 \end{center}
\end{figure}

In the literature, many authors have studied shock in accretion discs \citep{f87a,c89,cc11,ck16} and
the phenomena 
have long been identified as the result of centrifugal barrier developing in the
accreting flow. Shocks may develope in jets due to the interaction with the ambient medium,
or inherent fluctuation of injection speed of the jet.
But why would internal shock develop in steady jet flow, where the role of angular momentum
is either absent or very weak?
In Fig. \ref{lab:fig6}(a), 
we plot velocity profile of 
a shocked jet solution
with parameters 
$E=1.28$, $\xi=1.0$ and $x_{\rm sh}$=12. 
The jet, 
starting with subsonic speeds at base, 
passes through the sonic point (circle) 
at $r_c=5$ and becomes transonic. 
This sonic point is formed due to gravity 
term (third term in the r. h. s of equation \ref{dvdr.eq})
which is negative and equals other two terms 
at $r_c$ making the r. h. s. of 
equation (\ref{dvdr.eq}) equal to zero.
It is to be remembered that, jets with higher values of $E$ implies
hotter flow at the base, which ensures greater thermal driving which makes
the jet supersonic within few $\rg$ of the base. 
However, after the jet becomes supersonic ($v>a$), the jet accelerates but within a short distance
beyond the sonic point the jet decelerates (thin, solid line).
This reduction in jet speed occurs due to
the geometry of the flow. In Fig. (\ref{lab:fig6}g),
we have plotted the corresponding cross section of the jet.
The jet rapidly expands in the subsonic regime, but the expansion gets arrested and the expansion
of the jet geometry becomes very small ${\cal A}^{-1}d{\cal A}/dr\sim 0$. Therefore the positive contribution
in the r. h. s of equation (\ref{dvdr.eq}) reduces significantly which makes $dv/dr\leq 0$. Thus
the flow is decelerated resulting in higher pressure down stream
(thin solid curve of Fig. \ref{lab:fig6}d). This resistance causes the jet to under go
shock transition at $\rsh=8.21$. The shock condition is also satisfied at $\rsh=17.4$,
however this outer shock can be shown to be unstable \citep[see, Appendix \ref{shock_stability} and also][]{n96,yk95,ydl96}.
We now compare the shocked M2 jet in Fig. (\ref{lab:fig6}a, d, g) with two other jet flows,
(i) a jet of model M2 but with low energy $E=1.02$ (Fig. \ref{lab:fig6}b, e, h); and
(ii) a jet of model M1 and with the same energy $E=1.28$ (Fig. \ref{lab:fig6}c, f, i).
In the middle panels, $E=1.02$ and therefore the jet is much colder. Reduced thermal driving causes
the sonic point to form at large distance (open circle in Fig. \ref{lab:fig6}b).
The large variations in the fractional gradient of ${\cal A}$
occurs well within $r_c$. At $r>r_c$ ${\cal A}^{-1}d{\cal A}/dr \rightarrow~2/r$, which is
similar to a conical flow. Therefore, the r. h. s of equation (\ref{dvdr.eq}) does not become
negative at $r>r_c$. In other words, flow remains monotonic. The pressure is also a monotonic
function (Fig. \ref{lab:fig6}e) and therefore no shock transition occurs.
In order to complete the comparison, in the panels on the right
(Fig. \ref{lab:fig6}c, f, i),
we plot a jet model of M1, with
the same energy as the shocked one ($E=1.28$). Since fractional variation of
the cross section is monotonic i. e., at $r>r_c$, ${\cal A}^{-1}d{\cal A}/dr=2/r$ (Fig. \ref{lab:fig6}i),
the all jet variables like $v$ (Fig. \ref{lab:fig6}c) and pressure
(Fig. \ref{lab:fig6}f) remains monotonic. No internal shock develops.

Therefore to form such internal shocks in jets, the jet base has to be hot in order to make
it supersonic very close to the base. And then the fractional gradient of
the jet cross section needs to change rapidly, in order to alter the effect of gravity, so that the jet beam starts resisting the matter
following it and form a shock.

\begin{figure}
\begin{center}
 \includegraphics[width=11cm]{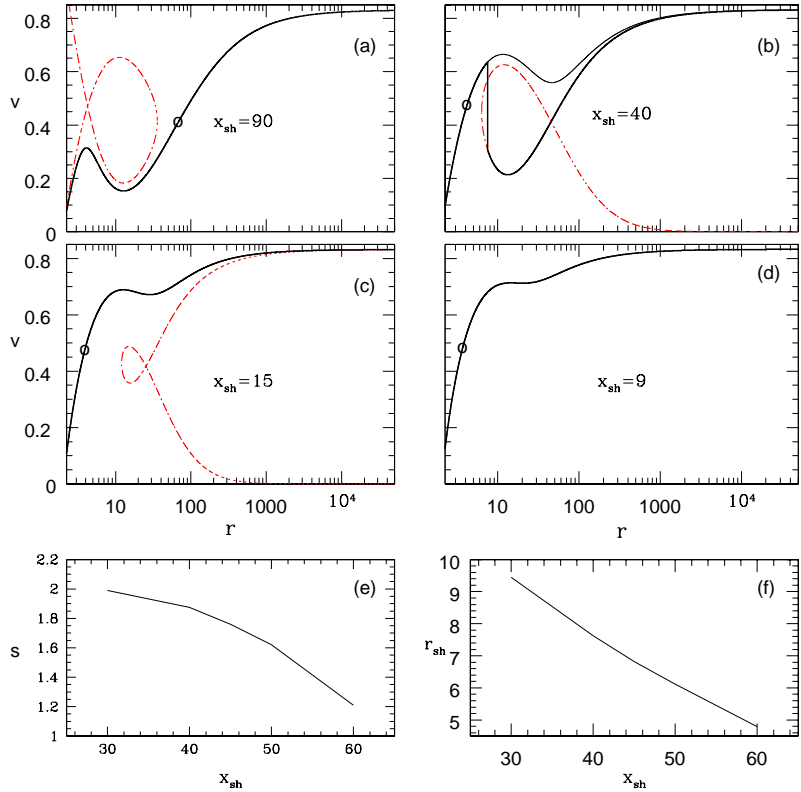}
\caption{Variation of $v$ along $r$ for different values of $x_{\rm sh}$ 
as labeled in (a)-(d). The dotted (red) curves are closed solutions,
while thin black solid curves are
possible transonic solutions. 
Vertical line show stable shock 
transition (Thick solid). Thick solid curves are the realistic 
transonic solutions of the jet;
(e) Shock strength $S$ as a function of  
$\xsh$; and (d) $\rsh$ 
as a function of $\xsh$. Here $E=1.8$, $\xi=1.0$.}
\label{lab:fig7}
 \end{center}
\end{figure}

Figures (\ref{lab:fig6}a-i) showed that departure of the jet
cross section from conical geometry is not enough to drive shock in jet.
It is necessary that the jet becomes transonic at a short distance form the base
and a significant fractional change in jet cross-section occurs in the supersonic regime
of the jet. Since the departure of the jet cross-section from the conical one, depends
on shape of the inner disc, or in other words, the location of the shock in accretion,
we study the effect of accretion disc shock on the jet solution.
We compare jet solutions (i. e., $v$ versus $r$)
for various accretion shock locations for e. g.,
$\xsh=90$ (Fig. \ref{lab:fig7}a), $\xsh=40$ (Fig. \ref{lab:fig7}b), $\xsh=15$
(Fig. \ref{lab:fig7}c) and $\xsh=9$ (Fig. \ref{lab:fig7}d).
In Figs. (\ref{lab:fig4}c-i), all possible solutions were obtained by keeping $\xsh$
constant for different values of $E$.
In Figs. (\ref{lab:fig7}a-d) we show how the jet solution changes for different values of
$\xsh$ but keeping $E=1.8$ of the jet same. For a large value of $\xsh=90$ (Fig. \ref{lab:fig7}a),
the jet cross section near the base
diverges so much that
the jet looses the forward thrust in the subsonic regime and the sonic
point is formed at large distance. The geometry indeed decelerates
the flow, but being in the subsonic regime such deceleration do not accumulate enough pressure to break the
kinetic energy and therefore no shock is formed. As the expansion of the cross-section is 
arrested, the jet starts to accelerate and eventually becomes transonic at large distance from the BH. At relatively smaller value of $\xsh~(=40)$, the thermal term remains strong enough
to negate gravity and form the sonic point in few $\rg$. For such values of $\xsh$, the fractional
expansion
of the jet cross-section drastically reduces or, ${\cal A}^{-1}d{\cal A}/dr\sim 0$,
when the jet is supersonic. Therefore, in this case
the jet suffers shock (Fig. \ref{lab:fig7}b). Infact, for $E=1.8$ the jet will under go shock transition, if the
accretion disc shock location range is from $\xsh=30$---$60$. 
For even smaller value of accretion shock location $\xsh=15$, because the opening angle of the jet is less,
the thermal driving is comparatively more than the previous case. The jet becomes supersonic at an even shorter distance. The outer sonic point is available, but because the shock condition
is not satisfied, shock does not form in the jet (Fig. \ref{lab:fig7}c).
As the shock in accretion is decreased to $\xsh=9$, the thermal driving is so strong that
it forms only one sonic point, overcoming the influence of the geometry (Fig. \ref{lab:fig7}d).
Although, due to the fractional change in jet geometry, the nature of jet solutions have changed, but
jets launched with same Bernoulli parameter achieves
the same terminal speed independent of any jet geometry. This is because at $r\rightarrow \infty$,
$h\rightarrow h_\infty \rightarrow 1$ $\Rightarrow ~u_t\rightarrow u_{t \infty}\rightarrow \gmt$, so  
$$
E=-hu_t=-h_{\infty}u_{t\infty}= \gmt=(1-v^2_\infty)^{-1/2}.
$$
or,
\begin{equation}
v_\infty=\left(1-\frac{1}{E^2}\right)^{1/2}	.
	\label{vterm.eq}
\end{equation}
In Figs. (\ref{lab:fig7}e, f), the jet shock strength (equation \ref{shokstrnth.eq}) and 
the jet shock location $\rsh$ are plotted as functions of accretion shock location $\xsh$.
  
\begin{figure}
\begin{center}
 \includegraphics[trim={0 0 0 2.5cm},clip,width=10cm]{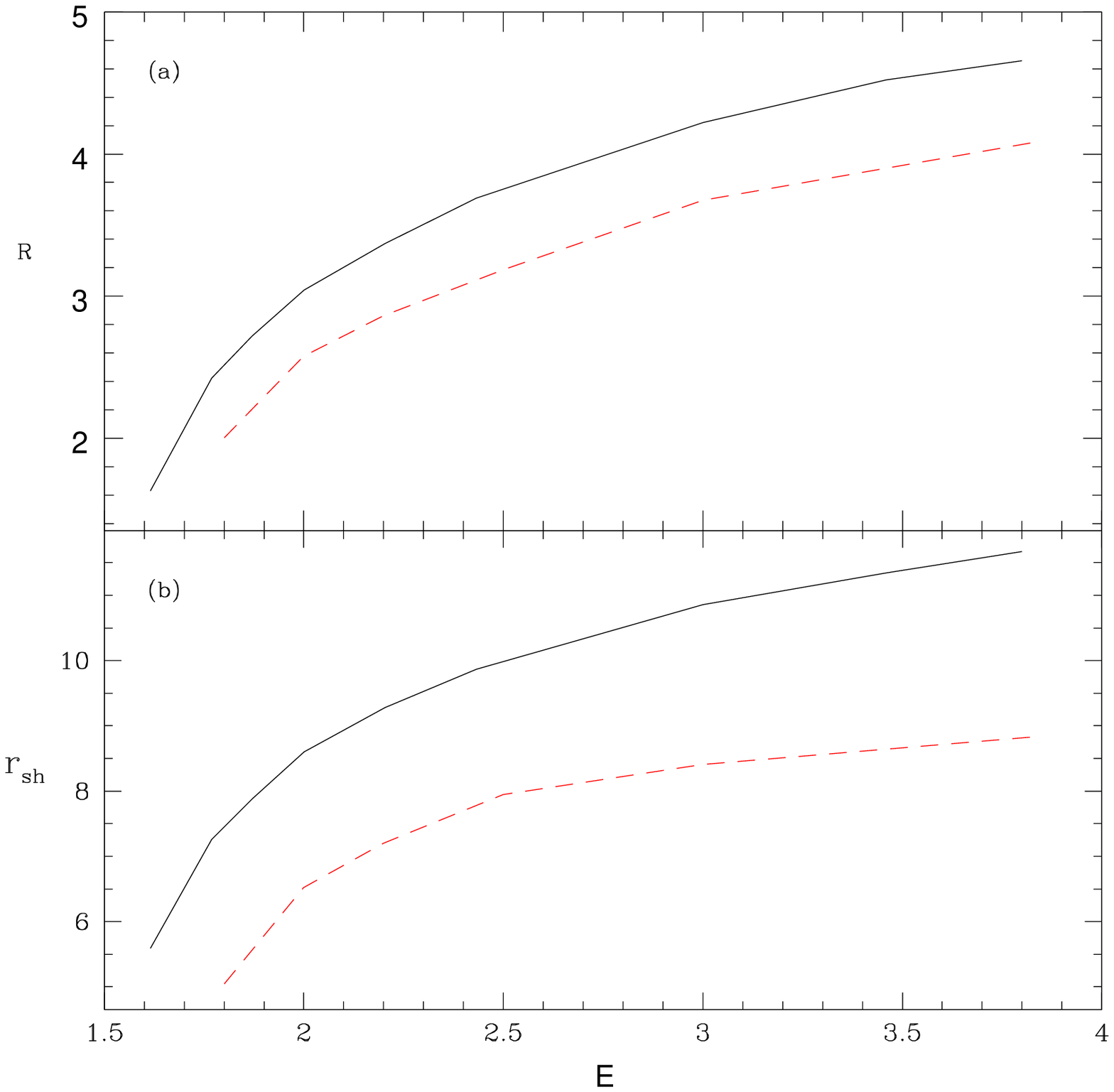}
 \caption{$R$ (a) and $\rsh$ (b) as functions 
 of $E$ for $\xsh=40$ (solid) and 
 $\xsh=60$ (dashed, red)}
\label{lab:fig8}
 \end{center}
\end{figure}
In Figs. (\ref{lab:fig8}a, b), the shock compression ratio $R$ (equation \ref{compress.eq}) of the jet and the
jet shock location $\rsh$ as a function of $E$ is plotted. Each curve represents the accretion shock location $\xsh=40$
(solid) and $\xsh=60$ (dashed). It shows that for a given $E$, the jet shock $\rsh$
and strength $S$ decrease
with the increase of $\xsh$. From Fig. (\ref{lab:fig4}a) it is also clear that for larger
values of $\xsh$,
jet shock may form in larger region of the parameter space.
The compression ratio of the
jet is above $3$ in a large part of the parameter space, therefore shock acceleration would be more efficient
at these shocks.
It is interesting to note the contrast in the behaviour of the jet shock $\rsh$ with the accretion disc shock.
In case of accretion discs, the shock strength and the compression ratio increases with decreasing shock location $\xsh$ \citep{kc13,ck16}. But for the shock in jet, the dependence of $R$ and $S$ on $\rsh$
is just the opposite i. e., $R$ and $S$ decreases with decreasing $\rsh$. 

\begin{figure}
\begin{center}
 \includegraphics[trim={0 0 0 6.3cm},clip,width=11cm]{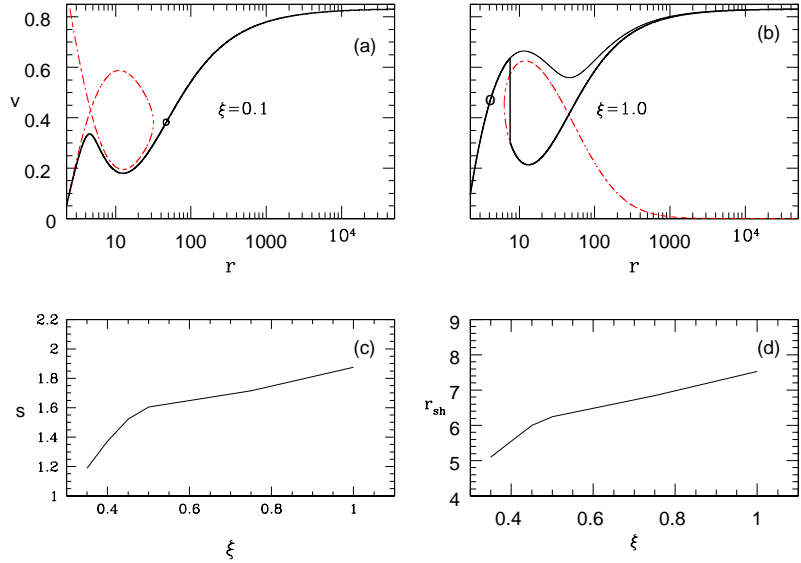}
 \caption{Variation of $v$ with $r$ for different values of $\xi=0.1$ (a) and $\xi=1.0$ (b), also
 labeled on the plots. Closed (dash) and global (solid) solutions. Thick  solid (black  online) curves show the transonic physical trajectory 
 of the jet and the vertical line is the shock;
(c) $S$ and (d) $\rsh$ 
with $\xi$ plotted. For all panels
$x_{\rm sh}=40$ and $E=1.8$.}
\label{lab:fig9}
 \end{center}
\end{figure}

So far we have only studied the jet properties for $\xi=1.0$ or, electron-proton ($\ep$) flow. In Newtonian flow,
composition would not influence the outcome of the solution if cooling is not present.
But for relativistic flow, composition
enters into the expression of the enthalpy and therefore, even in absence of cooling,
jet solutions depend on the composition. In Figs. (\ref{lab:fig9}a) we plot the velocity
profile of a jet whose composition corresponds to $\xi=0.1$ i. e., the proton number density is $10\%$
of the electron number density, where the charge neutrality is restored by the presence of positrons. 
The jet energy is $E=1.8$ and jet geometry is defined by $\xsh=40$. In Fig. (\ref{lab:fig9}b), we plot
the velocity profile of an $\ep$ jet ($\xi=1.0$), for the same values of $E$ and $\xsh$.
While $\ep$ jet undergoes shock transition, but for the same parameters $E$ and $\xsh$, for the flow with $\xi=0.1$ composition, there is no shock. Infact, $\xi=0.1$ jet solution is similar to the one with lower energy.
Although, the jet solutions of different $\xi$ are distinctly different at finite $r$, but the terminal
speeds are same as is predicted by equation (\ref{vterm.eq}). For these values of $E$ and $\xsh$,
shock in jets are obtained for $\xi=0.32 \rightarrow 1.0$.
In Fig. (\ref{lab:fig9}c) and (\ref{lab:fig9}d) 
 we plot shock strength $S$ and the shock location
 $\rsh$ respectively, as a function of $\xi$ for the same values of $E$ and $\xsh$. Shock
 produced in heavier jet ($\ep$) is 
 stronger and the shock is located at larger distance form the BH. Figures (\ref{lab:fig9}c-d) 
 again show that the shocks forming farther away from 
 BH, are stronger.  
 \begin{figure}
 	\begin{center}
 		\includegraphics[trim={0 0 0 2.8cm},clip,width=11cm]{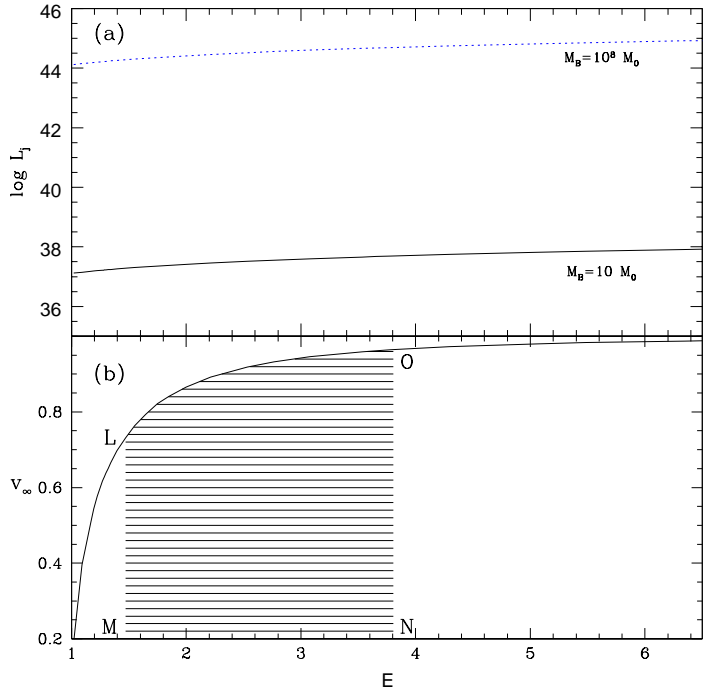}
 		\caption{(a)$v_\infty$  as a function of $E$. 
 			$\xsh$ and $\xi$ are kept 40 and 1.0 (b) $\log L_j$ with $E$ for $M_B=10^8 M_\odot$ 
 			(dot, blue) $M_B=10 M_\odot$ (solid, black)}
 		\label{lab:fig10}
 	\end{center}
 \end{figure}
 
 Although AGN jets are directly observed and superluminal motions of bright spots directly observed,
 but still in most cases the jet speeds and jet kinetic power are generally inferred.
 However, few decades of continuous multi wavelength monitoring of the objects have lend
 reasonable credibility in the inferred values of $\Lj$.
 From equations (\ref{ljet.eq}, \ref{vterm.eq}), it is quite clear that in absence of dissipation
 both $\Lj$ and $\vt$ are independent of jet geometry as well as, the composition of the flow.
 It must be remembered that, when expressed in terms of Eddington mass flow rate, $\Lj$ will depend on the
 central BH.
 In Fig. (\ref{lab:fig10}a), $\Lj$ is plotted as a function of $E$. The mass outflow rate is assumed to
 be ${\dot M}_{\rm out}=0.01{\dot M}_{\rm Edd}$.
 $\Lj$ scales with the mass of the BH and over a large range of $E$ and for a central BH of mass
 $10^8 M_{\odot}$, $\Lj$ ranges from $10^{44-45}$ ergs s$^{-1}$ (dotted), while for a $10M_\odot$ BH,
 the kinetic power of the jet is $10^{37-38}$ ergs s$^{-1}$. These estimate would go up or down depending on
 the accretion rate, as well as the mass of the compact object.
 In Fig. (\ref{lab:fig10}b), $\vt$ is plotted as a function of $E$. From equation (\ref{vterm.eq})
 it is clear that $\vt$ is a function of $E$ only, except, when there are other accelerating mechanism
 acting on the jet. So $\vt$ in this figure is true for both M1 and M2 jets. However, if we assume
 $\xsh=40$ for M2 jet and $\xi=1.0$, then the shaded region LMNOL corresponds to jets
 which undergo steady shock transitions.

It might be fruitful to explore, whether these internal shocks satisfies some observational features.
One may recall that the charged particles 
oscillates back and forth across a shock with horizontal width $L$
 and in each cycle, its energy keeps increasing. After successive oscillations,
 the particle escapes the shock region with enhanced  energy known as 
e-folding energy and is given by \citep{bi87}
\be
E_T=\frac{L(v_+-v_-)\bsh}{\pi c}\left(\frac{\rm erg}{\rm statC}\right),
\label{E_T.eq}
\ee
where $\bsh$ is the magnetic field at the shock and $L=2GM_Br_{\rm sh}\tan\theta/c^2$.
The typical magnetic field estimates near the horizon vary from $10mG$ \citep{l11} to 
$10^4G$ \citep{kl97}. For $M_B\sim 14\msol$(Cygnus X-1), the e-folding energy for the shock obtained in
Fig. \ref{lab:fig7}b is obtained to be $~16 MeV$-$1.6 TeV$ (per electron charge in statC) depending on different magnetic field estimates
mentioned above. The energy associated with high energy tail of $400keV-2MeV$ in Cygnus X-1 \citep{l11} can easily be explained by the internal jet shocks discussed in this paper. 
The spectral index of shock accelerated particles is 
 \be 
q=\frac{R}{R-1}
\ee
And for the same set of jet parameters we obtained $R=1.87$ or $q=2.15$,
therefore, even the estimated spectral index of $2.2 \pm 0.4$ of such observational estimates \citep{l11}
can be given a theoretical basis.

\section{Discussion and Concluding Remarks}
\label{sec5}

In this paper, we investigated the possibility of finding steady shocks in jets
very close to the compact object. Since the jets exhibit mildly relativistic to ultra relativistic
terminal speeds,
therefore, one has to describe the flow in the relativistic regime. The jets traverse massive
length scales originating very close to the central BH to distances more than hundred thousand
Schwarzschild radii, therefore gravitational effects need to be considered. And since relativity
and any form of Newtonian gravity is incompatible, therefore, the jet has to be described as
fluid flow in the general relativistic limit. In the present paper, we investigated
jets in Schwarzschild metric and have used a relativistic equation of state (as opposed to the Newtonian
polytropic one) to describe the thermodynamics
of the jet. Since jets are collimated and flows about the axis of symmetry, and as pointed out in section \ref{sec2}, the jet is generated with low angular momentum and it is likely to be further reduced
by other physical processes like viscosity, we consider non rotating jets as an approximation.

In this paper we have studied two jet models. The first model M1 was the conical jet or radial outflow
and the flow geometry is that of a cone. The second model M2 was assumed from physical argument and some
evidence found in previous jet simulations. Since the jet base is supposed to be very hot therefore, it is
expected that the jet would expand in all directions, only to be mechanically held by the funnel
shaped inner surface of the torus-like inner disc. The expansion of the jet cross-section
gets arrested above the inner torus like corona and then expands again at larger distances
from the BH. We assumed a jet flow geometry which follows this pattern.
It should also be remembered that, in this paper the accretion disc plays an auxiliary role.
We did not compute any
jet parameters from the accretion disc. We just used an approximate accretion disc solution
to define the flow geometry at the base of the jet, and then used the outer boundary of the
inner disc, or, $\xsh$ as a parameter which defines the departure of the jet geometry from
the conical cross-section. And to do that, we fitted the temperature distribution of the PSD
by an approximated function, which would determine the inner surface of the torus part of the inner disc or PSD. That inner surface was considered as the outer boundary of the jet
geometry close to the BH. The accretion disc solution used is exactly same as our previous paper \citet{vkmc15},
although unlike the previous paper, no other input from the accretion disc was used in the
jet solution. One must remember that accretion solutions change for different values of
disc parameters, although, in this paper the disc solutions do not influence jet solutions.

Since we have ignored other external accelerating mechanism or any dissipation, so $E$ is a constant of motion. An adiabatic jet is a fair assumption until and unless the jet interacts
with the ambient medium. Given these assumptions, terminal speed ($\vt$) of the jet is solely governed
by $E$ and would not depend on either the jet geometry, or, the composition of the flow.
However, the jet geometry influences the solutions at finite $r$ from the BH. While M1 model
showed monotonic smooth solutions, whose terminal speed increased
with the increasing $E$.  But for M2 model, depending on the jet energy $E$ and the
accretion disc parameter $\xsh$, we obtained very fast, smooth jets flowing out through one
inner X-type sonic point; and for other combinations of $E~\&~\xsh$, we obtained jets with multiple sonic points and even shocks. For very low $E$ of course, weak jets flow out though the outer X-type sonic point.
In connection to Fig. \ref{lab:fig6} in section \ref{sec:m2}, it has been clearly explained that,
in order to obtain steady state internal shocks, the jet material in the supersonic regime
has to oppose the flow following it. That can happen if the expansion of the
flow geometry drastically decrease when the jet is in the supersonic regime, making gravity
relatively stronger. The effect would not be so effective at distances were gravity is itself weak. Therefore the necessary condition is that, the expansion of the jet geometry
decreases drastically in few to few$\times ~10\rg$. Along with this, $E$ has to be high enough,
so that the enhanced thermal driving makes the flow supersonic before the region 
where ${\cal A}^{-1}d{\cal A}/dr \sim 0$. One of the advantages of a shock driven by 
the jet geometry is that, it has no impact on either the kinetic power
of the jet or, its terminal speed. So if $E$ is high enough produce very strong jet with
relativistic terminal speed, then in addition one may have shock jump in the jet without compromising the $\vt$.
It should be understood, that in presence of other accelerating mechanism, $E$ will not
remain constant and would increase outwards. Therefore, the terminal speeds obtained here may be considered
as the minimum that can be achieved for the given input parameters.
The jet kinetic power (equation \ref{ljet.eq}) obtained also depends on ${\dot M}$ and $E$, so output
would depend on the central mass and the mass supply. In our estimation of $L_{\rm j}$ we assumed
higher mass outflow rate at the base, but general relativistic estimates of such mass loss is around few percent
of the accretion rate, which can easily explain the estimates for Cygnus X-1 jet \citep{rfgk07}

Interestingly, for low values of $\xsh$, the jet geometry of M2  differs slightly from the conical one.
Therefore, the jet shock is obtained in
a very small range of $E$. For higher $\xsh$, the range of $E$ which can harbour jet shocks
also increases. However, at same $E$, the jet shock $\rsh$ is formed at larger distance from the BH, if $\xsh$ is formed closer to the BH. This is very interesting, because in accretion disc
as the shock moves closer to the BH, the shock becomes stronger. In addition, smaller value of $\xsh$
implies higher values of $\rsh$ and higher $\rsh$ means stronger jet shock. That means a very hot gas may surround the BH, not only
in the equatorial plane but also around the axis as well. For the current objectives 
the detailed spectral analysis is beyond the scope of the paper and has 
not been done for such a scenario as yet, we would like to do that in future.  

The assumptions made in this paper, were to simplify the jet problem and study all possible relativistic
jet solutions possible in the presence of strong gravity. However, few things
definitely would influence the conclusions of this paper positively. Consideration of radiation driving for intense radiation field would surely affect the jet solution. It would surely
accelerate the jets \citep[see,][]{vkmc15}, but in addition radiation drag in presence of an intense and isotropic radiation field may drive shocks, which means even M1 jet may harbour shocks. Moreover, the jet geometry need not be conical at
large distances, and the jet geometry may be pinched off by other mechanisms, which might create
more shocks. Furthermore, entrainment of the jet at large distances might qualitatively modify the conclusions of this paper. Even then, the result of this paper might be important in many other ways. From Figs. (\ref{lab:fig6}, \ref{lab:fig7}, \ref{lab:fig8}) it is clear that, 
farther the shock forms, stronger is the shock. Since the base of the jet is much hotter
than even the post-shocked region, then any surge in mass outflow rate, or, energy/momentum enhancement in the base will try to drive the jet away from its stable position. If this
driving is too strong, it will render the shock unstable and the shock would travel out along the jet beam.
This should have two effects, the unshocked jet material will be shock heated, moreover,
since by the nature of jet shocks, it becomes stronger for larger $\rsh$, therefore a shock
traveling out would shock-accelerate jet particles quite significantly. And this would happen
comparatively close to the jet base. However, if the driving is not too strong,
this might lead to quasi periodic oscillations. Such shock oscillation model has not been
properly probed even in micro-quasar scenario. This might even be interesting for QPOs in
blazars. It must be remembered that, previously, we did not obtain standing internal shocks in jets around
non-rotating BHs \citep{kc13,kc14,ck16} for geometries assumed. The main aim of this paper is to show that, if jet geometries with non-spherical geometry is considered, then internal shocks may form
close to the jet base and such solutions may address few observational features as well.
And the assumptions made in this paper was aimed at reducing the frills, which might obfuscate the real driver
of such shocks in jets.

Drawing concrete conclusions, one may say, M2 jets with energies $E=1.5$---$2.5~c^2$ may harbour
shocks in the range $\rsh = 5.5$---$10 \rg$ from the central BH. The shocks are in the range of
compression
ratio $R=1.6$---$4.0$. The terminal speeds of these jets are between $\vt=0.745$---$0.916c$. 

\section*{Acknowledgment}%
The authors acknowledge the anonymous referee for helpful suggestions to improve the quality of this paper.

{}

\appendix
\section{Stability analysis of the shocks}
\label{shock_stability}
In section \ref{results}, we reported existence of 2 shocks, 
present at either sides of middle sonic point in various 
solutions. One of the two such shocks were shown to be unstable previously
\citep{n96,yk95,ydl96}. We include the stability analysis for the sake of
completeness.
The momentum flux, $T^{rr}$ (equation \ref{sk3.eq}), 
remains conserved across the shock. But if the shock 
under some perturbation, moves 
from $r_{\rm sh}$ to $r_{\rm sh}+\delta r$ then $T^{rr}$ may 
not be balanced. The resultant difference across the shock is 
\be 
\delta T^{rr}=T^{rr}_2-T^{rr}_1=\left[\left(\frac{dT^{rr}}{dr}\right)_2-\left(\frac{dT^{rr}}{dr}\right)_1\right]\delta r = \Delta \delta r
\label{delta.eq}
\ee
Multiplying and dividing equation (\ref{sk3.eq}) 
by $\rho$ and after rearranging the expression for 
momentum flux becomes 
\be 
T^{rr}=\rho \left(hu^ru^r+\frac{2\Theta g^{rr}}{\tau}\right)
\label{Trr2.eq}
\ee
Now using equation (\ref{mdotout.eq}) and 
differentiating equation (\ref{Trr2.eq}) 
followed by some algebra, one obtains
\be 
\frac{dT^{rr}}{dr}=\frac{{\dot M_{out}}}{{\cal A \tau}}\left[-A_s \frac{dv}{dr}+B_s\right],
\label{Trr3.eq}
\ee
where, 
\be
A_s={\sqrt{g^{rr}} \gamma^3}\left[ \frac{2 \Theta g^{rr}}{u^2}+\left(f+2 \Theta \right) \right]+\frac{2 \Theta {\gamma}^2}{Nuv}\left(u^2(N+1)+g^{rr}\right)
\label{as.eq}
\ee
and
\begin{eqnarray}
B_s=\frac{\gamma v}{r^2 \sqrt{g^{rr}}}\left[ \left(f+2 \Theta \right)-\frac{2 \Theta g^{rr}}{u^2} \right]-\left((f+2\Theta) u^2+2 \Theta g^{rr}\right)\frac{1}{u {\cal A}}\frac{d{\cal A}}{dr} \\ \nonumber
+\frac{4 \Theta}{ur^2}-\frac{2 \Theta}{N u}\left[u^2(N+1)+g^{rr}\right]
\label{bs.eq}
\end{eqnarray}
Using equation (\ref{Trr3.eq}) in (\ref{delta.eq}), we obtain
\be 
\Delta=\left[-A_s v'+B_s\right]=\left(A_{s1}v'_1-A_{s2}v'_2\right)+(B_{s2}-B_{s1})
\label{delta2.eq}
\ee
Now, the stability of the shock depends on 
the sign of $\Delta$. If $\Delta<0$ for finite 
and small $\delta r$ there is more momentum flux 
flowing out of the shock than the flux flowing in 
so the shock keeps shifting towards further 
increasing $r$, and is unstable. On 
the other hand if $\Delta>0$, then the change 
due to $\delta r$ leads to the further decrease 
in $\Delta$, and the shock is stable.

One finds that equations (\ref{as.eq}) and (\ref{bs.eq}) 
$A_s$ has positive value. Now 
the stability of the shock can be analyzed under two 
broad conditions.\\
$\bullet$ Condition 1. The shock is significantly away 
from middle sonic point, or the absolute magnitude of 
$v'$ is significantly more than 0. We find that 
$|B_s|<<|A_s|$ and  hence the stability of the 
shock depends upon the sign of $v'$. Equation 
(\ref{delta.eq}) shows that the shock is stable 
(or $\Delta > 0$) if $v'_1>0$ and subsequently 
$v'_2 < 0 $.  Hence the inner shock is stable 
and the outer shock is unstable.\\
$\bullet$ Condition 2. If the shock is close 
to the middle sonic point, $v'_1 \approx v'_2 \approx 0$.
So only second term consisting $B_s$ contributes to the 
stability analysis and one obtains that $\Delta<0$ for 
both inner and outer shocks and the shock is always 
unstable.\\

Finally, the general 
rule for stability of the shock is, \\
$\bullet$ If the post shock flow is accelerated then 
the shock is unstable and if the post shock flow is 
decelerated the shock is stable unless the shock is 
very close to the middle sonic point where it is 
unstable.\\ 

In the paper all the stable shocks are shown by
thick solid vertical lines and 
unstable shocks are represented by thin dotted 
vertical lines.

\section{Approximated accretion disc quantities}
\label{disc_structure}
The jet geometry of M2 model was taken according to
equation (\ref{geomvar.eq}). The inner part of the accretion disc or PSD shapes the jet
geometry near the base.
\begin{figure}
\begin{center}
 \includegraphics[trim={0 0 0 6.5cm},clip,width=11cm]{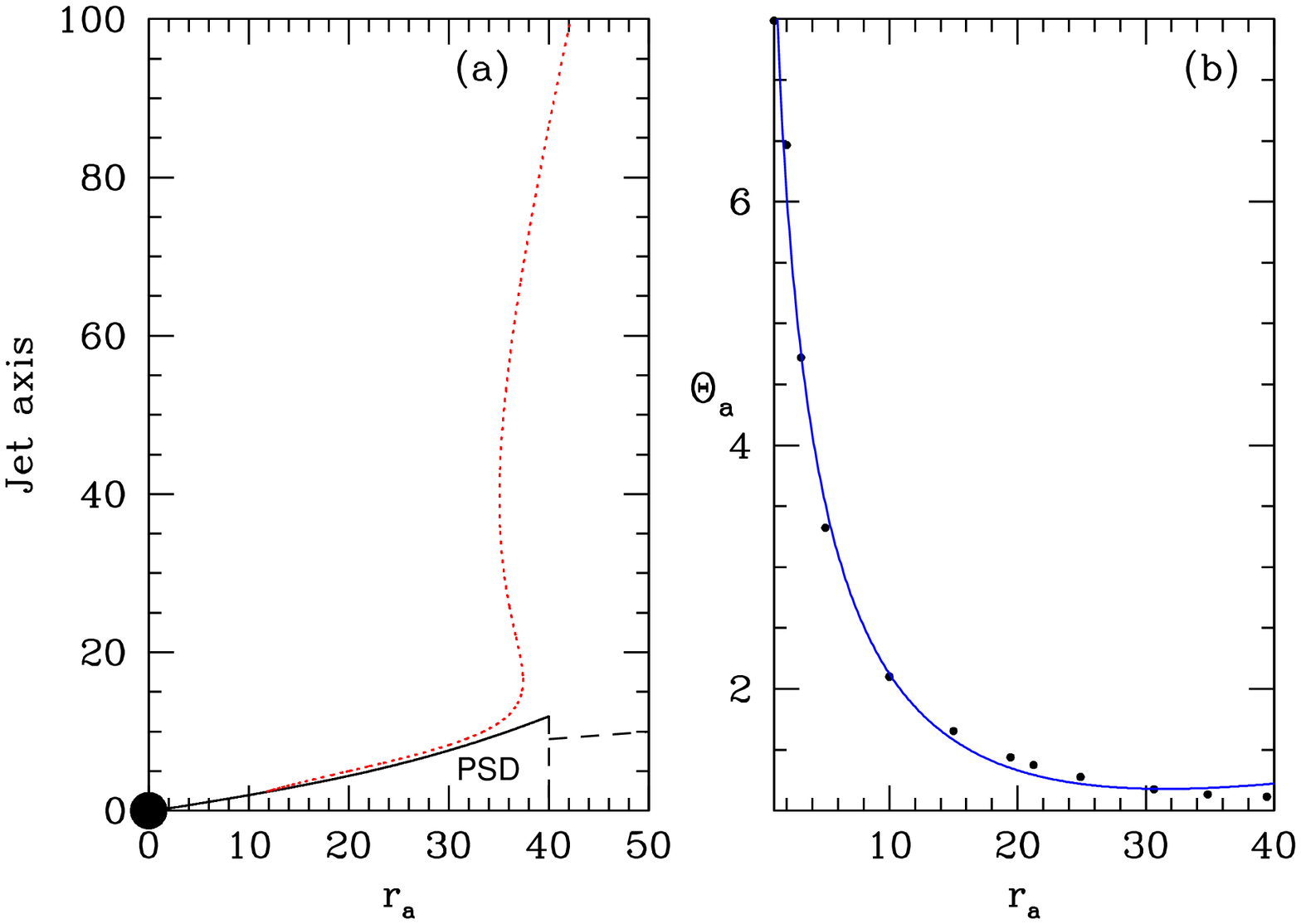}
 \caption{(a) Accretion disc height $H_{\rm a}$ is plotted with $r_{\rm a}$ with solid line (black online). Shock location is at $x_{\rm sh}=r_{\rm a}=40$, shown by long dashed line. The jet width $r sin \theta$ is over plotted with red dotted line. Black hole resides at $r_{\rm a}=0$ with shown as black sphere with $r_s=2$ (b) Fitted $\Theta_{\rm a}$.
 }
\label{lab:accretion_disc_jet_plot}
 \end{center}
\end{figure}
If the local density, four velocity, pressure and dimensionless temperature in the accretion disc are $p_{\rm a}$, $u^r_{\rm a}$, $\rho_{\rm a}$ and $\Theta_{\rm a}$, respectively, and the angular momentum of the disc is $\lambda$, then the local height $H_{\rm a}$ of the post shock region for a advective disc is given by \citep{cc11}
\be 
H_{\rm a}=\sqrt{\frac{p_{\rm a}}{\rho_{\rm a}}\left[r_{\rm a}^3-{\lambda}^2(r_{\rm a}-2)\right]}= \sqrt{\frac{2 {\Theta}_{\rm a}}{\tau}\left[r_{\rm a}^3-{\lambda}^2(r_{\rm a}-2)\right]},
\label{h_a.eq}
\ee
where, $r_{\rm a}$ is the equatorial distance from the black hole.
We obtain $\Theta_{\rm a}$ in an approximate way following \citet{vkmc15}.
The $u^r_{\rm a}$ is obtained by solving geodesic equation.
Since $u^r_{\rm a}$ is known at every $r_{\rm a}$, and accretion rate is a constant, so $\rho_{\rm a}$
is known. We also know that $\Theta_{\rm a}$ and $\rho_{\rm a}$ are related by the adiabatic relation. So
supplying $\Theta_{\rm a}$, $\rho_{\rm a}$ and $u^r_{\rm a}$ at the $\xsh$, we know $\Theta_{\rm a}$
for all values of $r_{\rm a}$. We plot the $\Theta_{\rm a}$  in Fig. (\ref{lab:accretion_disc_jet_plot}b)
for $\xsh=40$ in filled dots.
We obtain an analytic function to the variation of $\Theta_{\rm a}$ with $r_{\rm a}$.
The obtained best fit is 
\be
\Theta_{\rm a} = \exp(-r_{\rm a}^{a_t}+{b_t})c_t+d_tr.
\label{Theta_a.eq}
\ee
Here, $a_t=0.391623$, $b_t=2.30554$, $c_t=2.22486$ and $d_t=0.0225265$. 
This fit is shown in Fig. (\ref{lab:accretion_disc_jet_plot}-b) by solid line with 
points being actual values of $\Theta_{\rm a}$.
The fitted function is used to compute $H_{\rm a}$ and
is plotted in Fig.  (\ref{lab:accretion_disc_jet_plot}-a) with solid line. We over plot the jet structure (equation \ref{geomvar.eq}) in red dots. 

 \end{document}